\documentclass[nonacm,acmtog,screen]{acmart}
\setcopyright{cc}

\usepackage{multirow}
\usepackage{color, colortbl}
\usepackage{subcaption}
\usepackage{graphicx}
\usepackage[utf8]{inputenc}

\usepackage[T1]{fontenc}
\usepackage{csquotes}
\usepackage{microtype}
\usepackage{times}
\usepackage{latexsym}
\usepackage{amsmath}
\usepackage{float}
\usepackage{footnote}
\usepackage{enumitem}
\usepackage{bm}
\usepackage{arydshln}
\usepackage{listings}
\usepackage{booktabs}
\usepackage{multicol}
\usepackage{xcolor}
\usepackage{bbding}
\usepackage{makecell}
\usepackage{mathtools}
\usepackage{imakeidx}
\makeindex
\usepackage{lipsum}
\usepackage{natbib}
\usepackage[toc]{multitoc}
\usepackage[edges]{forest}
\usepackage[normalem]{ulem}

\setcopyright{none}

\definecolor{mydarkblue}{rgb}{0,0.08,0.45}
\definecolor{openaigreen}{RGB}{85, 180, 129}

\usepackage[most]{tcolorbox}
\usepackage{soul}
\newtcolorbox{prompt}{
  colback=openaigreen!15,
  colframe=gray!60,
  boxrule=1pt,
  arc=0pt,
  boxsep=0pt,
  left=6pt,
  right=6pt,
  top=6pt,
  bottom=6pt,
  enhanced,
  fontupper=\small,
  grow to left by=-1mm,
  grow to right by=-1mm,
}

\begin{document}

\title[Evaluating Higher Education Vulnerability to AI Assistants]{Could ChatGPT get an Engineering Degree? Evaluating Higher Education Vulnerability to AI Assistants}

\author{Beatriz Borges}
\authornote{Authors B.B., N.F., D.B., and A.S. contributed equally to this research.}
\author{Negar Foroutan}
\authornotemark[1]
\author{Deniz Bayazit}
\authornotemark[1]
\author{Anna Sotnikova}
\authornotemark[1]
\author{Syrielle Montariol}
\author{Tanya Nazaretzky}
\author{Mohammadreza Banaei}
\author{Alireza Sakhaeirad}
\author{Philippe Servant}
\author{Seyed Parsa Neshaei}
\author{Jibril Frej}
\author{Angelika Romanou}
\author{Gail Weiss}
\author{Sepideh Mamooler}
\author{Zeming Chen}
\author{Simin Fan}
\author{Silin Gao}
\author{Mete Ismayilzada}
\author{Debjit Paul}
\author{Philippe Schwaller}
\author{Sacha Friedli}
\author{Patrick Jermann}
\author{Tanja K\"aser}
\author{Antoine Bosselut}
\authornote{Corresponding author: antoine.bosselut@epfl.ch}
\affiliation{%
\institution{EPFL}
\city{Lausanne}
\country{Switzerland}}

\author{EPFL Grader Consortium}
\authornote{Alexandre Schöpfer, Andrej Janchevski, Anja Tiede, Clarence Linden, Emanuele Troiani, Francesco Salvi, Freya Behrens, Giacomo Orsi, Giovanni Piccioli, Hadrien Sevel, Louis Coulon, Manuela Pineros-Rodriguez, Marin Bonnassies, Pierre Hellich, Puck van Gerwen, Sankalp Gambhir, Solal Pirelli, Thomas Blanchard, Timothée Callens, Toni Abi Aoun, Yannick Calvino Alonso, Yuri Cho}

\author{EPFL Data Consortium}
\authornote{Aleksandra Radenovic, Alexandre Alahi, Alexander Mathis, Anne-Florence Bitbol, Boi Faltings, Cécile Hébert, Devis Tuia, François Maréchal, George Candea, Giuseppe Carleo, Jean-Cédric Chappelier, Nicolas Flammarion, Jean-Marie Fürbringer, Jean-Philippe Pellet, Karl Aberer, Lenka Zdeborová, Marcel Salathé, Martin Jaggi, Martin Rajman, Mathias Payer, Matthieu Wyart, Michael Gastpar, Michele Ceriotti, Ola Svensson, Olivier Lévêque, Paolo Ienne, Rachid Guerraoui, Robert West, Sanidhya Kashyap, Valerio Piazza, Viesturs Simanis, Viktor Kuncak, Volkan Cevher, Akhil Arora, Alberto Chiappa, Antonio Sclocchi, Étienne Bruno, Florian Hofhammer, Gabriel Pescia, Geovani Rizk, Leello Dadi, Lucas Stoffl, Manoel Horta Ribeiro, Matthieu Bovel, Yueyang Pan}

\renewcommand{\shortauthors}{Borges, et al.}

\begin{abstract}
AI assistants are being increasingly used by students enrolled in higher education institutions. While these tools provide opportunities for improved teaching and education, they also pose significant challenges for assessment and learning outcomes. We conceptualize these challenges through the lens of \textit{vulnerability}, the potential for university assessments and learning outcomes to be impacted by student use of generative AI. We investigate the potential scale of this vulnerability by measuring the degree to which AI assistants can complete assessment questions in standard university-level STEM courses. Specifically, we compile a novel dataset of textual assessment questions from 50 courses at EPFL and evaluate whether two AI assistants, GPT-3.5 and GPT-4 can adequately answer these questions. We use eight prompting strategies to produce responses and find that GPT-4 answers an average of 65.8\% of questions correctly, 
and can even produce the correct answer across at least one prompting strategy for 85.1\% of questions. When grouping courses in our dataset by degree program, these systems already pass non-project assessments of large numbers of core courses in various degree programs, posing risks to higher education accreditation that will be amplified as these models improve. Our results call for revising program-level assessment design in higher education in light of advances in generative AI.
\end{abstract}


\maketitle

\section{Introduction}
ChatGPT, a system using a large language model~(LLM), GPT-3.5, as its foundation, was released in November 2022 to broad adoption and fanfare, reaching 100M users in its first month of use and remaining in the public discourse to this day. As arguably the most hyped AI system to date, its release has prompted a continuing discussion of societal transformations likely to be induced by AI over the next years and decades. Potential changes in modern educational systems have remained a core topic in this discussion, with early reports highlighting the risk of these AI systems as tools that would allow students to succeed in university coursework without learning the relevant skills those courses are meant to teach. Despite this concern, there has yet to be a comprehensive empirical study of the potential impact of LLMs (and derivative tools) on the assessment methods that educational institutions use to evaluate students. A few studies have explored the interesting sub-task of how well models perform on problems related to topics typically taught in many university courses and aggregated relevant question sets for this purpose \citep{hendrycks2021measuring, huang2023ceval, wang2023scibench, zhong2023agieval, arora2023llms}. However, none of these works extrapolate these findings to assess the downstream impact of these tools on degree programs, where the risk of these technologies relative to their pedagogical benefits must actually be measured. 

\begin{figure}
    \centering
    \includegraphics[width=.8\linewidth]{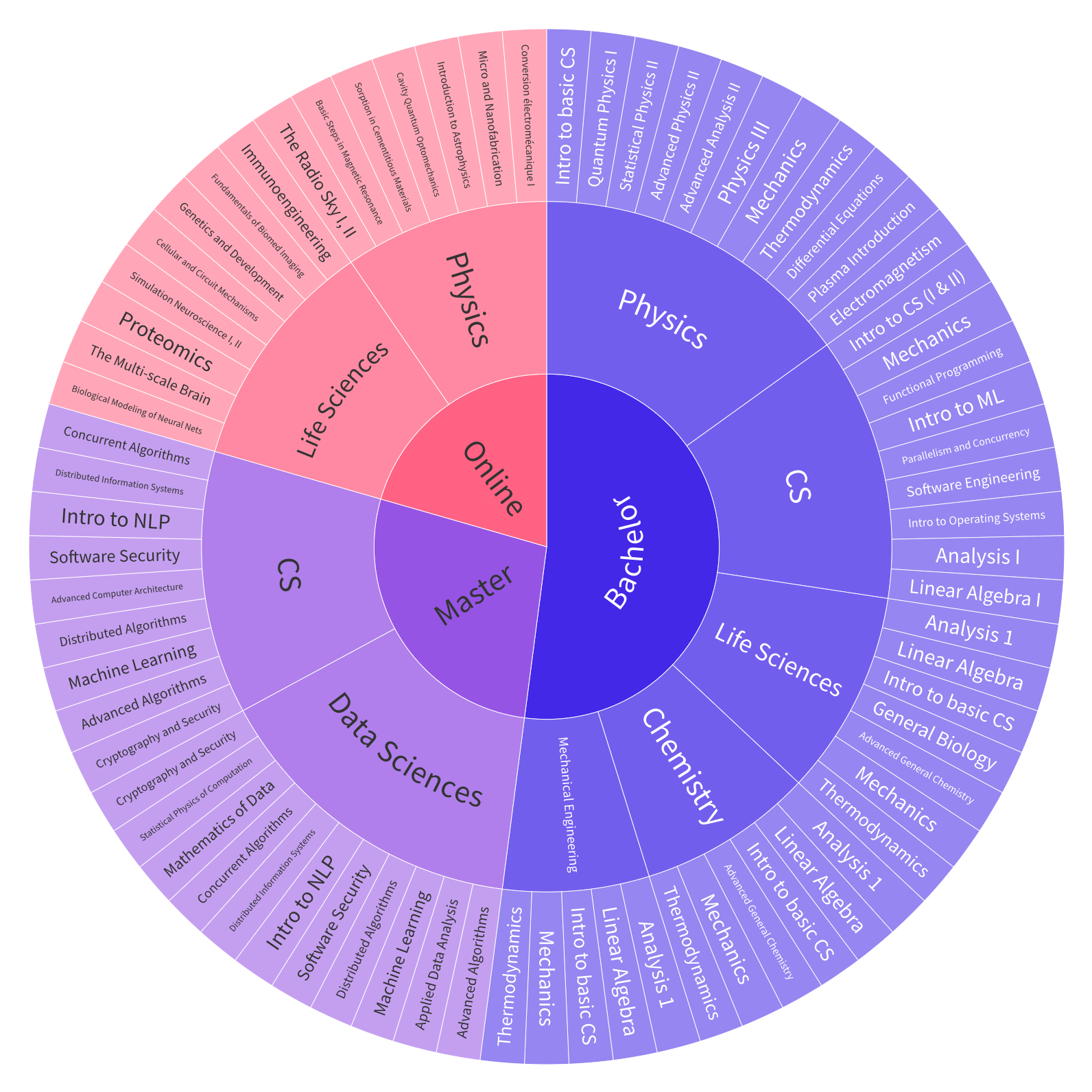}
    \caption{\textbf{Overview of Courses.} Courses represented in our dataset, grouped by program and degree. Courses may belong to multiple programs, in which case their partition is split into chunks of equal size, with one chunk assigned to each program.}
    \label{fig:courses}
\end{figure}

In this work, we conduct a large-scale study across 50 courses from EPFL to measure the current performance of LLMs on higher education course assessments. The selected courses are sampled from 9 Bachelor's, Master's, and Online programs, covering between 33\% and 66\% of the required courses in these programs. From these courses, we compile a bilingual dataset (English and French) of 5,579 textual open-answer and multiple-choice questions~(MCQ). All questions were extracted from real exams, assignments, and practical exercise sessions used to evaluate students in previous years. The course distribution is presented in Figure \ref{fig:courses}, and the dataset statistics are shown in Table \ref{tab:data} (stratified by particular dataset attributes). 

Using this dataset, we subsequently test two commonly-used models, GPT-4 \citep{openai2023gpt4}, the system widely considered to be the most performant \citep{zheng2024judging} among public AI assistants,\footnote{as of November 2023} and GPT-3.5 \citep{openai2022gpt3.5}, a highly performant freely available system. We generate responses from these systems using a range of prompting strategies \citep{brown2020language,xu2023expertprompting,wei2023chainofthought,yao2023tree,wang2023selfcritique,madaan2023selfrefine,wang2023metacognitive}, each of which varies in complexity, but all of which could easily be applied by a lay practitioner with minimal training in prompt engineering \citep{Sahoo2024ASS}. We evaluate these systems using both automatic and manual grading, where manual grading of open-answer questions is performed by the same faculty staff that designed these problems and who have experience grading student answers to them. Subsequently, we conduct a detailed analysis of the generated outputs and their assessment results, considering factors such as the number of courses that would be passed, their distribution across university programs, as well as the effects of the question difficulty and language.

Our results show that AI systems are relatively capable of answering questions used in university assessments. GPT-4 responds correctly to $\sim$65.8\% of questions when aggregating responses across the different prompting strategies using a simple majority vote (i.e., a \textit{knowledge-free} setting that assumes the user would use this tool with no subject knowledge). Furthermore, across the eight prompting strategies, GPT-4 can generate at least one correct response for 85.1\% of questions (maximum performance), indicating even greater assessment vulnerability in a setting where a user may have enough subject knowledge to \textit{select} a {correct} answer even if they cannot produce it. This performance is relatively stable across courses in various scientific disciplines, impacting courses regardless of their subject and size. Importantly, we find that these results indicate that large numbers of university degree programs are highly vulnerable to these tools, with the non-project components of many core courses being passed in multiple degrees offered by our institution.

Finally, we observe that while these systems are capable of reaching passing grades in many university assessments, they struggle with more complex question types where students also tend to perform most poorly. Taken together, these results indicate a possibility that these systems could be used to achieve passing marks in university courses while circumventing the process by which students acquire basic domain knowledge and learn to extend it to more complex problems. We conclude with a discussion on mitigations to university assessment settings, an outlook on how university systems should adapt to the increased use of these tools, and a discussion of limitations of our study, specifically with respect to how it diverges from exact assessment and grading policies at our institution.
%Finally, we observe that while these systems are capable of reaching passing grades in many university assessments, they struggle with more complex question types where students also tend to perform most poorly. Taken together, these results indicate a possibility that these systems could be used to achieve passing marks in university courses while circumventing the process by which students acquire basic domain knowledge and learn to extend it to more complex problems. We conclude with a discussion on short-term mitigation to university assessment settings, as well as an outlook on how university systems should adapt to the increased use of these tools.

\begin{table}[t]
\centering
\footnotesize
\begin{tabular}{c l r}
\toprule
&\textbf{Category} & \textbf{Total Questions}\\
\midrule
\multirow{3}{*}{\textbf{Level}} & Bachelor's courses &  2,147 (38.5\%)\\
  & Master's courses & 1,631 (29.2\%)\\
  & Online programs & 1,801 (32.3\%)\\
 \midrule
\multirow{2}{*}{\textbf{Language}} & English &  3,933 (70.5\%)\\
  & French & 1,646 (29.5\%)\\
 \midrule
\multirow{2}{*}{\textbf{Question Type}} & MCQ &  3,460 (62\%)\\
  & Open-Answer & 2,119 (38\%)\\
\bottomrule
\end{tabular}
\vspace{2pt}
\caption{\textbf{Dataset Statistics.}}\label{tab:data}
\end{table}

\section{Data Collection}  
We compile a new dataset of assessment questions from 50 courses offered at our institution from both on-campus and online classes. Following data pre-processing and filtering steps, this dataset consists of a total bank of  5,579 textual multiple-choice (MCQ) and open-answer questions in both English and French. These questions span various levels (e.g., Bachelor, Master), and cover a broad spectrum of STEM disciplines, including Computer Science, Mathematics, Biology, Chemistry, Physics, and Material Sciences. Table \ref{tab:data} and Figure \ref{fig:courses} provide an overview of the dataset's main statistics and the distribution of questions across different topics. Additionally, we have collected course-specific attributes such as the year when the course is first offered in our institution's degree programs (e.g., \textit{Master's year 1}), the program designation (e.g., \textit{Physics}), the language of instruction (e.g., \textit{French}), and the average student enrollment over recent years. Finally, certain questions have been labeled by the instructor who designed the question with a subjective annotation of the question's difficulty.

\section{Experimental Findings}
In our study, we investigate the vulnerability of university programs to generative AI systems using our question bank of 5,579 evaluation questions from 50 courses. We consider two models, GPT-4 and GPT-3.5, selected due to their popularity and usage rates. GPT-4 is considered the most performant model among all publicly accessible LLMs but is only available through a premium subscription, impeding its use by many students. GPT-3.5 is a less performant alternative, but free to use. We generate responses to questions from these models using eight relatively easy-to-apply prompting methods (implementation details are described in Appendix \ref{ap:prompt}). For multiple-choice questions, we assess whether a response is correct by comparing the selected choice with the annotated correct answer option. For open-response questions, we use GPT-4 to rate the quality of the response on a four-point scale: \texttt{Correct, Mostly Correct, Mostly Incorrect, Incorrect}, which we map to scores of 1, 0.66, 0.33, and 0, respectively, for calculating performance.\footnote{Analysis of the quality of this automated grading is provided in Appendix \ref{ap:grading_eval}. Importantly, we note that GPT-4 gives slightly higher average grades (on average $\sim$2.75\%) than humans for responses to a sample of questions graded by both.}

% Appendix \ref{ap:additional_results}) \autoref{ap:grading_eval}.

\begin{figure}
    \centering
    \includegraphics[width=1\linewidth]{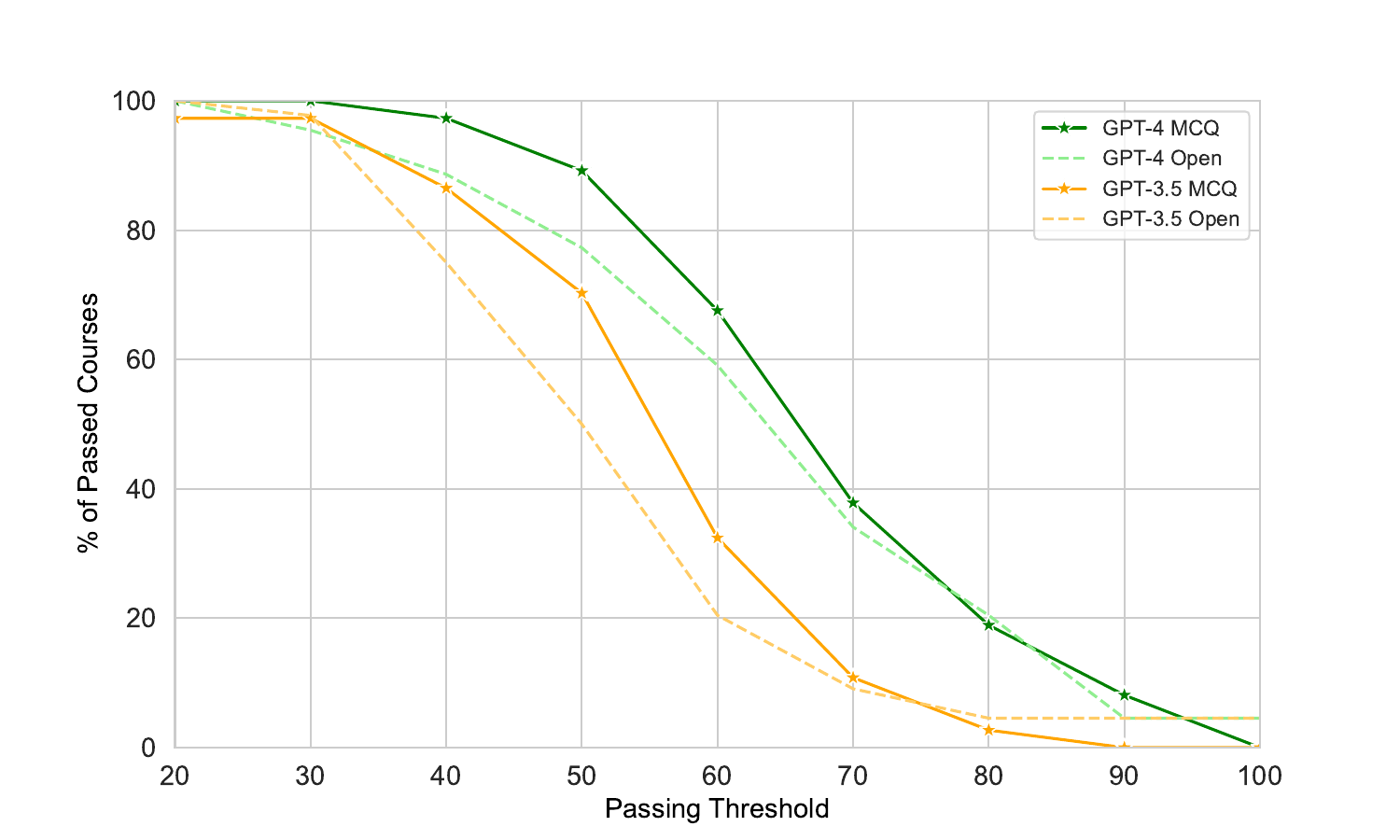}
    \caption{\textbf{Course Pass Rate of Generative AI Assistants.} Proportion of 50 courses that models pass at various performance thresholds. Results are presented independently for multiple-choice (MCQ) and open-answer (Open) question types for both GPT-3.5 and GPT-4. Model responses are aggregated using the majority vote strategy.}
    \label{fig:course-passing-rate}
\end{figure}

\subsection{Can LLM systems pass university-level courses?}
We begin our analysis by assessing model performance in a setting where the user has zero knowledge about the question topic. {In the simplest scenarios, where we use the most straightforward prompting strategies, such as directly asking a question (zero-shot) or asking the model to provide a reasoning chain before answering the question (chain-of-thought zero-shot), GPT-4 achieves average accuracies of 55.9\% and 65.1\%, respectively.} However, if the user employed a slightly more complex zero-knowledge strategy, such as majority voting over the eight answers generated by the different prompting strategies, they would receive a correct answer to 65.8\% (on average) of questions using GPT-4 (and 52.2\% using GPT-3.5).  We observe that this performance trend holds regardless of the language of the assessment, with English exhibiting slightly better results than French. Further experimental results for assessments in English and French are detailed in Appendix \ref{ap:language}.

Beyond overall performance across the question bank, Figure \ref{fig:course-passing-rate} presents the proportion of passed courses for our sample of 50 courses based on varying passing thresholds. Alarmingly, GPT-4 can easily be used to reach a 50\% performance threshold (which could be sufficient to pass many courses at various universities) for 89\% of courses with MCQ-based evaluations and for 77\% of courses for open-answer questions. While not performing quite as well, GPT-3.5, the freely available model, can reach a 50\% threshold for 70\% of courses with MCQ-based assessments and for 50\% of courses with open-answer questions. As passing thresholds may not be set to 50\% for all institutions, we vary this threshold and find that GPT-4 still passes 68\% of courses at a 60\% passing threshold, and 38\% of courses at a 70\% passing threshold for MCQ. Similar results are found for open-answer questions. 

Our results suggest that users with no knowledge of a particular subject could solve enough assessment questions to pass a majority of the courses in our dataset, making a compelling argument that AI assistants could potentially augment student learning as support tools. However, they simultaneously present a credible short-term risk to educational systems if these systems are not adapted to protect against the misuse of these technologies. Finally, we expect this problem to only grow worse over time, as continual model improvements in the years to come will make these tools even more performant in academic contexts.

\begin{table}[t!]
\resizebox{\linewidth}{!}{
\begin{tabular}{lccccc}
\toprule
\multirow{2}{*}{\textbf{Program}} & \multicolumn{3}{c}{\textbf{\% Courses Passed}} & 
       \multirow{2}{*}{\textbf{Max}} & \multirow{2}{*}{\parbox{1.3cm}{\textbf{Question}\\ \textbf{Count}}}\\
       & $\tau$=$50\%$ & $\tau$=$60\%$ & $\tau$=$70\%$ & &\\
\midrule
Engineering & 80.0 & 60.0 & 40.0 & 0.83&1,343\\
Chemistry &83.3  &66.7& 50.0 &0.85&1,417\\
Life Science & 85.7 &71.4&57.1 &0.85&1,477\\
\midrule
Physics Bachelor & 100.0& 55.6& 33.3&0.86 & 958\\
CS Bachelor & 91.7 & 66.7&50.0 &0.87 &1,487\\
\midrule
CS Master & 100.0& 83.3&50.0 & 0.87& 1,514\\
Data Science Master & 90.0 & 70.0&30.0 &0.86 & 1,576\\
\midrule
Physics Online & 100.0& 63.6&27.3 &0.84 &837\\
Life Science Online & 85.7&71.4&57.1&0.75&996\\
\bottomrule
\end{tabular}
}
\vspace{2pt}
\caption{\textbf{Program results.} For each program, the first three columns show the percentage of courses for which GPT-4 surpasses the thresholds $\tau$ = 50, 60, 70\% correctly-answered questions using the majority vote strategy. ``Max'' represents the proportion of questions in this degree correctly answered by at least one prompting strategy. Program levels are specified as Bachelor, Master, or Online. Engineering, Chemistry, and Life Science are first-year Bachelor programs.}\label{tab:program-stat}
\end{table}

\subsection{How do these results affect university programs?} The average performance across courses demonstrates each course's potential vulnerability to generative AI tools, which is particularly important if considerable portions of degree programs can be completed using these tools. To evaluate this program's vulnerability, we aggregate the questions in our datasets according to the study programs in which they are core courses. Specifically, we include four program types: 1st-year Bachelor, Full Bachelor, Full Master, and Online courses. We separate the first year of the Bachelor's degree because, at many institutions (including ours), the first year of the Bachelor's may have a fairly standardized curriculum that serves a special purpose (e.g., replacing or complementing entrance exams).

Full Bachelor's and Master's correspond to regular Bachelor's and Master's programs. We also include online courses, as official certifications can often be awarded for completing a sequence of these courses. For each program, our dataset contains a sample of courses that cover from 33\% to 66\% of the required courses for that program. For more program statistics, see Appendix \ref{ap:program_stat}.
%see Appendix \ref{ap:program_stat}.

We consider the same two aggregation strategies across the responses provided by the eight prompting methods: majority vote and maximum performance. For the majority vote, given the eight prompting strategies we have, the final answer to the question is the one that is the most frequent across all strategies. In the maximum performance aggregation, only a single prompting strategy is required to answer correctly for the model to be deemed correct in its response, approximating a pseudo-oracle setting that remains contextually realistic, as a user might be able to distinguish the answer when presented with it, even if they could not find it on their own. 

\begin{figure*}
 \centering
 \begin{subfigure}[t]{.32\textwidth}
     \centering
     \includegraphics[width=\textwidth]{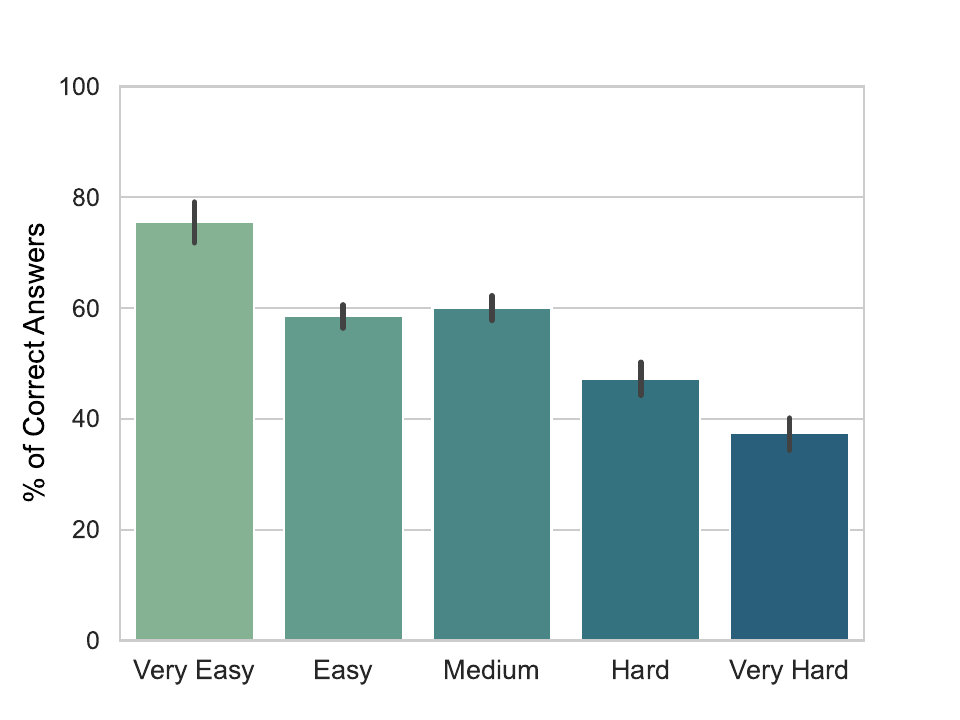}
    \caption{Question Difficulty Bachelor}\label{fig:diff}
 \end{subfigure}
\begin{subfigure}[t]{.32\textwidth}
     \centering
     \includegraphics[width=\textwidth]{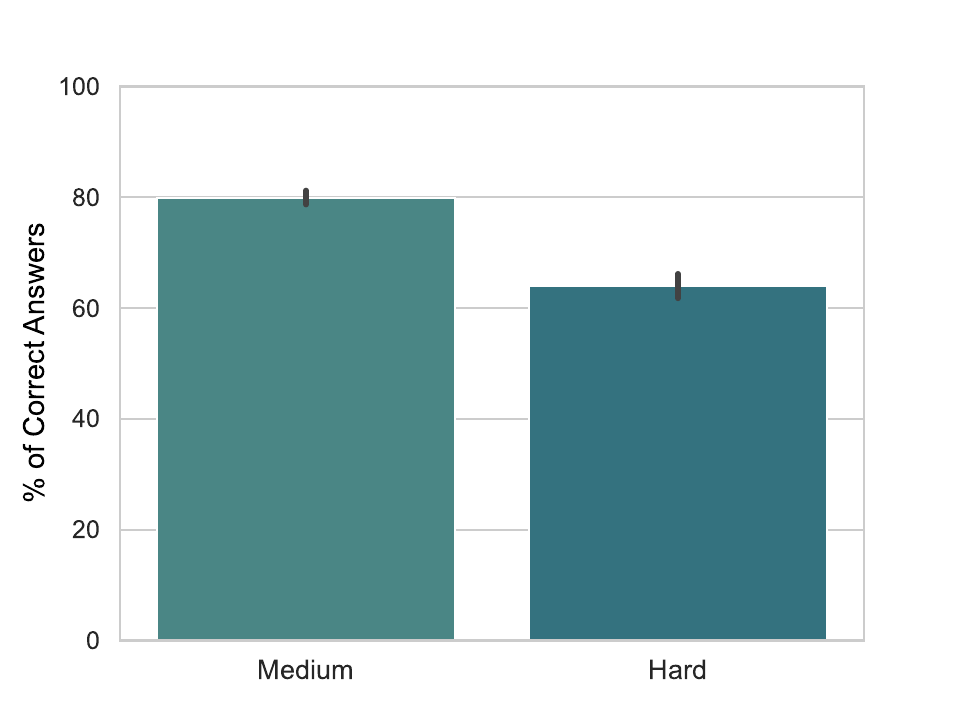}
   \caption{Question Difficulty Master}\label{fig:diff_2} \end{subfigure}
    \begin{subfigure}[t]{.32\textwidth}
     \centering
     \includegraphics[width=\textwidth]{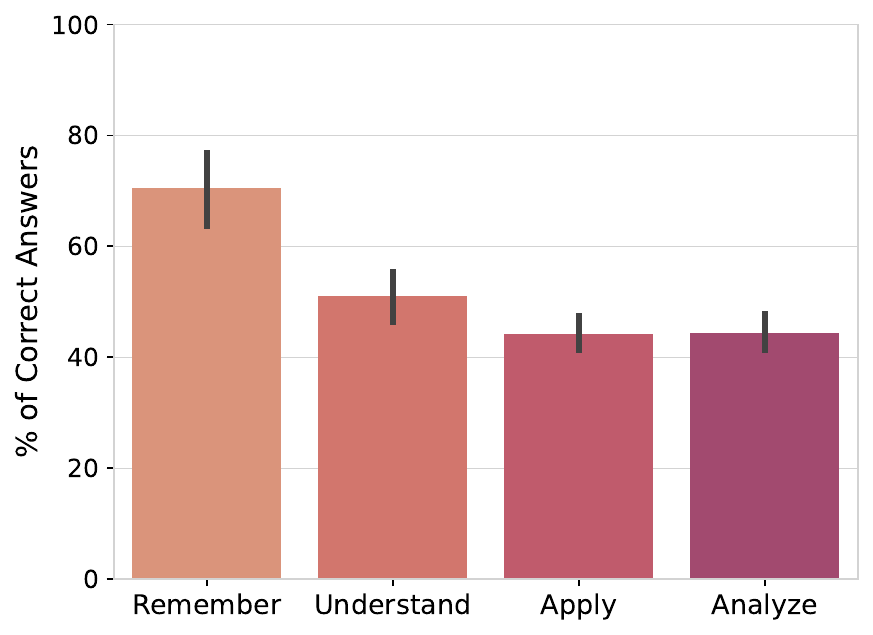}
     \caption{Bloom's Taxonomy}\label{fig:bloom}
 \end{subfigure}
\caption{\textbf{Model Performance Stratified by Question Difficulty.} (a, b) 376 Bachelor's and 693 Master's questions, respectively, annotated using instructor-reported difficulty levels. (c) 207 questions annotated using Bloom's taxonomy by two researchers in the learning sciences. Across all categorization schemes, GPT-4 performance slightly degrades as the questions become more complex and challenging. Performance is aggregated by the majority vote strategy. Error bars represent 95\% confidence intervals using the non-parametric bootstrap with 1000 resamples.}\label{fig:allignment}
 \end{figure*}

In Table \ref{tab:program-stat}, we present the number of courses that would be passed by GPT-4 across the 9 tested degree programs for various course passing thresholds $\tau$ (i.e., the performance that must be achieved to consider the course passed). Our results show that the general course vulnerability observed in the previous section extends to program vulnerability. At the $\tau = 50\%$ threshold for passing a course, at least 83\% of courses are passed in each of the evaluated programs. In certain programs, such as the Physics Bachelor and Computer Science Master, all tested courses are passed. While this number drops as we raise the passing threshold $\tau$, the maximum performance for each program typically remains above 80\%, indicating that a combination of clever prompting and partial subject knowledge may be sufficient to achieve high marks on assessment questions. 

Topically, we find that the models consistently exhibit lower performance on assessments of courses involving mathematical derivations. Conversely, the model demonstrates strong performance on problems that require generating code of text. For example, models consistently yield high performance in subjects such as \textit{Software Engineering} and \textit{Intro to Machine Learning}. This observation is further supported by the difference in performance between Master's and Bachelor's level courses (shown across Figures \ref{fig:diff} and \ref{fig:diff_2}). In our dataset, Bachelor's courses feature more mathematical derivations, while Master's courses have more text and code-based problems. In Appendix \ref{ap:course}, we provide further performance comparisons between the courses representing each program. In Appendix \ref{ap:prompting_strategy}, we analyze model performance across all prompting strategies and both question types.

% \begin{figure*}
%  \centering
%  \begin{subfigure}[t]{.32\textwidth}
%      \centering
%      \includegraphics[width=\textwidth]{figures/new_figures/Diff_bac.pdf}
%     \caption{Question Difficulty Bachelor}\label{fig:diff}
%  \end{subfigure}
% \begin{subfigure}[t]{.32\textwidth}
%      \centering
%      \includegraphics[width=\textwidth]{figures/new_figures/Diff_mas.pdf}
%    \caption{Question Difficulty Master}\label{fig:diff_2} \end{subfigure}
%     \begin{subfigure}[t]{.32\textwidth}
%      \centering
%      \includegraphics[width=\textwidth]{figures/new_figures/Bloom.pdf}
%      \caption{Bloom's Taxonomy}\label{fig:bloom}
%  \end{subfigure}
% \caption{\textbf{Model Performance Stratified by Question Difficulty.} (a, b) 376 Bachelor's and 693 Master's questions, respectively, annotated using instructor-reported difficulty levels. (c) 207 questions annotated using Bloom's taxonomy by two researchers in the learning sciences. Across all categorization schemes, GPT-4 performance slightly degrades as the questions become more complex and challenging. Performance is aggregated by the majority vote strategy. Error bars represent 95\% confidence intervals using the non-parametric bootstrap with 1000 resamples.}\label{fig:allignment}
%  \end{figure*}

\begin{figure}
    \centering
    \includegraphics[width=.9\linewidth]{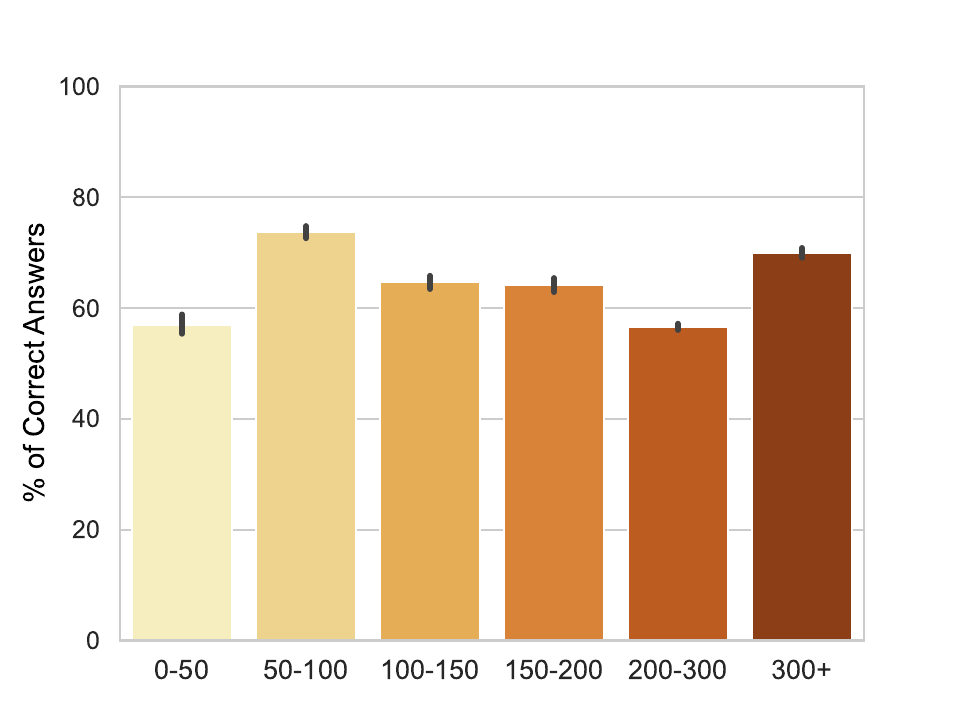}
    \caption{\textbf{Course Performance by Course Size.} Average course performance of GPT-4 with the majority vote strategy stratified by the course size, measured by the number of enrolled students. GPT-4 successfully answers questions for assessments in some of the largest courses by enrollment, amplifying the potential impact of assessment vulnerability. Error bars represent 95\% confidence intervals using the non-parametric bootstrap with 1000 resamples.}\label{fig:course-size}
\end{figure}

% \b

Finally, we highlight that these models are effective in courses that large portions of the student body must take, increasing the overall vulnerability of course programs. Figure \ref{fig:course-size} demonstrates that some of the largest classes on campus, with upwards of 300 students, are also some of the most vulnerable, with GPT-4 achieving (using the majority vote strategy) an average performance of 69.9\% in these classes (while hovering around 60\% for other class sizes). This result is particularly problematic because larger courses are often limited in terms of the monitoring and mitigation strategies they can implement due to the number of students. While smaller courses may more easily be able to combat the misuse and unethical use of generative AI tools, larger courses, which are often mandatory for degree completion, must ensure fair and scalable solutions for a larger student population.

% \begin{figure}[t]
%     \centering
% \begin{minipage}{.49\textwidth}
% \centering
%     \includegraphics[width=.9\linewidth]{figures/new_figures/Size.pdf}
%     \caption{\textbf{Course Performance by Course Size.} Average course performance of GPT-4 with the majority vote strategy stratified by the course size, measured by the number of enrolled students. GPT-4 successfully answers questions for assessments in some of the largest courses by enrollment, amplifying the potential impact of assessment vulnerability. Error bars represent 95\% confidence intervals using the non-parametric bootstrap with 1000 resamples.}\label{fig:course-size}
% \end{minipage}
% \hfill
% \begin{minipage}{.49\textwidth}
%     \centering
%     \vspace{8pt}
%     \includegraphics[width=\linewidth]{figures/new_figures/Student_Model_perf_3.pdf}
%     % \vspace{1pt}
%     \caption{\textbf{Comparison of student performance and GPT-4.} Average student performance for a subset of 197 questions is computed and stratified along 10-point intervals from 0 to 100. The model's performance with the majority vote strategy is assessed by human graders using a 4-point scale. We observe the model typically answers correctly questions that students also excel at. However, there are questions on which the model struggles, but students perform reasonably well.}\label{fig:student-model}
% \end{minipage}
% \end{figure}

\subsection{More challenging assessments are a half-solution.} 
One possible solution to mitigate assessment vulnerability would be to increase their difficulty beyond what generative AI systems are capable of solving, as we observe that the performance of these systems does decrease on more challenging assessment questions (Figure \ref{fig:allignment}). We measure the difficulty using a sub-sample of our question bank that is annotated according two different categorizations of their difficulty: (1) \textit{instructor-reported question difficulty}, a five-point difficulty categorization for Bachelor courses and two-point for Master's courses provided by the course instructors, and (2) \textit{Bloom's taxonomy} \citep{bloom1956taxonomy}, a six-point scale that measures the cognitive complexity of a question.\footnote{More details about Bloom's Taxonomy can be found in Appendix \ref{ap:bloom}.}

For the instructor-reported difficulty categorization, we collect annotations from course instructors for a subset of $376$ questions from the Bachelor's program (\textit{n.b.}, the instructors that designed the questions). The instructor-reported rating ranges from ``Very Easy'' to ``Very Hard'' on a 5-point scale. We also collect $693$ questions from the Master's program annotated on a 2-point scale, ranging from ``Medium'' to ``Hard'' (the original scale was meant to be 3-point, but no instructor reported an ``Easy'' question). In Figures \ref{fig:diff} and \ref{fig:diff_2}, we show the model's performance on questions stratified by their difficulty rating and observe that GPT-4 performs worse on questions that instructors deem harder. For example, in Bachelor courses, there is a 38\% difference in accuracy between ``Very Easy'' and ``Very Hard'' questions. However, the differences between Bachelor's ``Easy'' and ``Hard'' questions or Master's ``Medium'' and ``Hard'' questions are only 11.5\% and 15.75\%, respectively.

This pattern is also observed in our assessment of question difficulty performed using Bloom's taxonomy, which classifies educational learning objectives into levels of complexity and specificity: remember, understand, apply, analyze, evaluate, and create. Two researchers in the learning sciences manually annotated 207 questions from our dataset according to Bloom's taxonomy \citep{bloom1956taxonomy}. While the taxonomy typically associates questions into six categories, we found that most course assessment questions were covered by the first four categories. The results, grouped by taxonomy category in Figure \ref{fig:bloom}, illustrate that GPT-4 performance is negatively correlated with the cognitive complexity of the question, with higher performance on lower-level tasks compared to analysis and application tasks. 

\begin{figure}
    \centering
    \includegraphics[width=\linewidth]{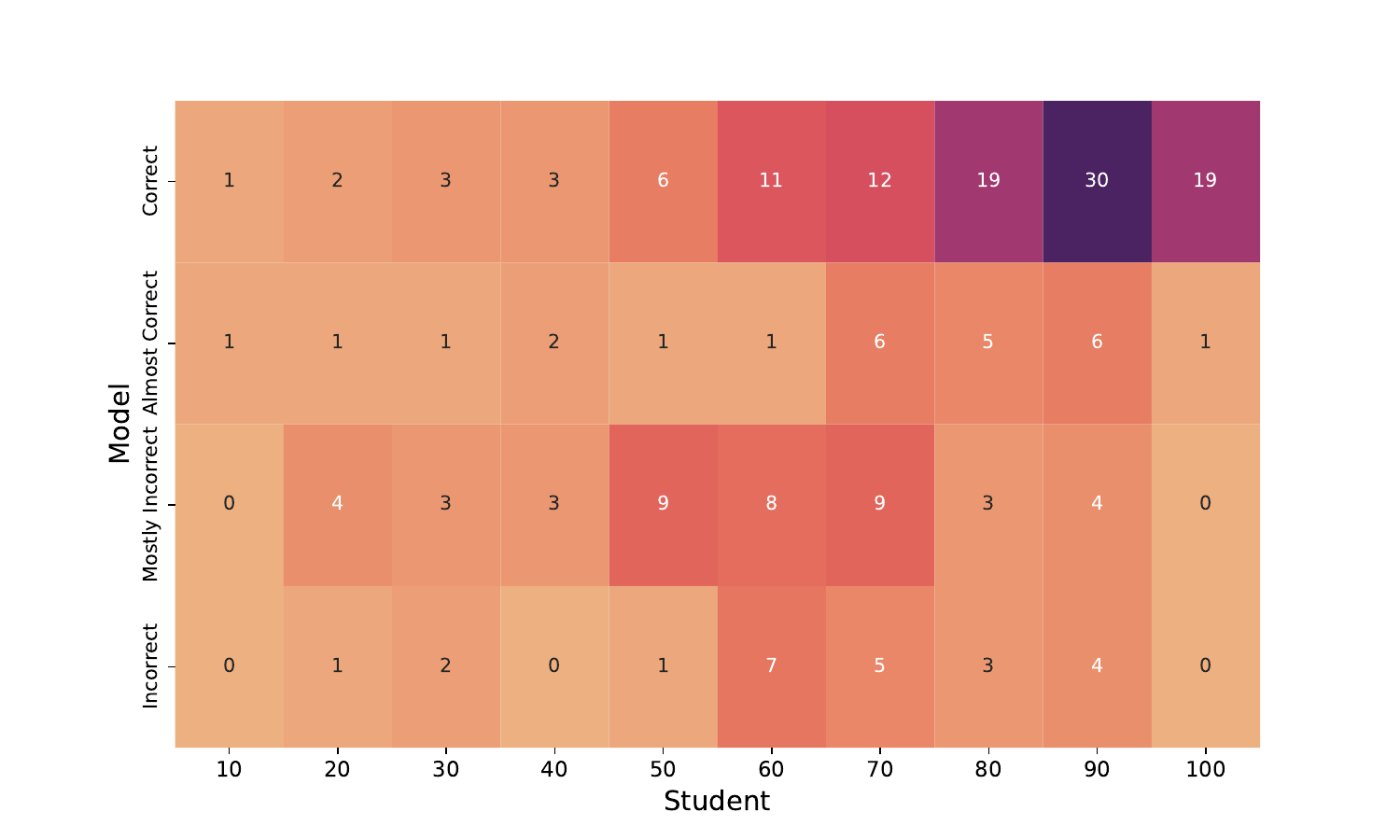}
    \caption{\textbf{Comparison of student performance and GPT-4.} Average student performance for a subset of 197 questions is computed and stratified along 10-point intervals from 0 to 100. The model's performance with the majority vote strategy is assessed by human graders using a 4-point scale. We observe the model typically answers correctly questions that students also excel at. However, there are questions on which the model struggles, but students perform reasonably well.}\label{fig:student-model}
\end{figure}

However, harder assessments may not necessarily be a suitable solution for this vulnerability as they would also likely lead to lower student performance. When comparing the performance of students and GPT-4 on problem sets from a subset of questions for which student performance statistics were collected (Figure \ref{fig:student-model}), we note that the model tends to excel on questions where students also perform well. This pattern perhaps exacerbates fairness as GPT-4 (and similar models) could be used to achieve average results on an assessment while providing few benefits to students who can already complete the easier portion of assessments but struggle with harder questions. Notably, however, we observe that for a subset of problems, the model typically struggles, receiving ``Mostly Incorrect'' or ``Incorrect'' marks, while students demonstrate relatively strong performance, with scores ranging from 0.5 to 0.9. These problems typically require mathematical derivations and extensive computations.

\section{Discussion}

\textbf{Summary.} In this work, we tested the ability of LLMs to solve assessment questions for a large number of courses from technical and natural sciences academic programs at EPFL. We find that LLMs are generally capable of answering 50-70\% (depending on the model) of questions correctly given no subject-related knowledge, and up to 85.1\% of questions correctly when some subject-specific knowledge is assumed (i.e., the ability to recognize the correct answer). Most importantly, when considering performance across programs, GPT-4 can achieve greater than 50\% performance for 83\% to 100\% of courses (depending on the program) with an average program pass rate of 91.7\%. While a higher per-course passing threshold of 70\% would only result in 23\% to 50\% of courses being passed across our programs (with an average of 37\%), this would also lead to higher student fail rates as passing thresholds would similarly affect them. Given that continuous advancements in LLM technology will likely further improve these LLM performance numbers in the future, we conclude that higher-education assessment schemes are immediately vulnerable to student exploitation of generative AI tools, specifically in the engineering and natural sciences.

\vspace{2mm}
\noindent \textbf{Assessment Vulnerability.} Our results indicate that the broad availability of generative AI tools should urgently trigger discussion on the design and implementation of assessments. Naturally, our results must be placed in the context where they would normally be observed. In many educational institutions, student assessments are frequently closed-book, thereby precluding the direct use of generative AI tools. Many course assessments (e.g., assignments), though, are completed at home without supervision. In the same vein, most online courses typically evaluate students without any supervised, in-person assessment. In these unsupervised settings, the availability of generative AI tools for aiding in the completion of assessments poses risks for many commonly used student evaluation methods. 

One general trend that we observe from our data (Figures \ref{fig:diff}, \ref{fig:diff_2}, and \ref{fig:bloom}) is that models perform well on basic learning tasks, such as memorizing factual knowledge. In courses where memorization of factual knowledge is a core evaluation objective, students should not be allowed to use these tools in non-proctored settings, and these courses should perhaps return to traditional in-person examination~\citep{Wang_2023}. Barring this possibility, in the case of non-proctored assessments, we recommend that their design should not only consider the possibility of assistance from generative AI but actively assume its use. At the very least, assessments should be designed with generative AI in the loop to design AI-adversarial evaluation that remain fair to students. 
 
At the same time, these findings provide an opportunity to improve and diversify student learning through assessments. Students acquire relevant skills when assessments emphasize analytical and applied knowledge settings \citep{zhang2023study}. Rather than using proctored exams, then, or limited practical works, such as assignments, students should be evaluated using assessments requiring a more composite application of course concepts, such as larger class projects. Project settings more closely assess students on problems resembling real-world challenges, would provide students with more opportunities to make problem-solving decisions, such as problem simplification, decomposition, and planning~\citep{Montgomery}, and would mitigate the impact of generative AI tools (Figure~\ref{fig:bloom}).

\vspace{2mm}
\noindent \textbf{Education Vulnerability.} While our results point to an urgent need to mitigate assessment vulnerabilities in higher education, a longer-term view requires considering how education as a practice should evolve alongside the availability of generative AI tools. Since the release of ChatGPT, ongoing discussions have revolved around this topic with both negative and optimistic views. While numerous studies explore the ways AI can enhance learning, ethical concerns related to plagiarism, biases, and overreliance on technology have also been highlighted~\citep{Yan_2023, chen2023generative, Pinto, ALQAHTANI20231236, Currie, DWIVEDI2023102642, lan2024survey}.

An important dimension of these discussions emphasizes the need to revisit evaluation procedures to ensure that students acquire necessary skills and critical thinking abilities in the face of AI integration~\citep{Prather_2023, becker2022programming, Finnie-Ansley, nowrozy2024embracing}. For instance, observations from various works (and our study) show that models excel in generating code to solve software problems~\citep{vaithilingam2022expectation, xu2022systematic, li2022competition, hou2024systematic, liu2024your}. While this capability reduces the burden on professional (and hobbyist) software developers, it also poses a risk for learners by potentially offering shortcuts that impede the acquisition of fundamental coding and engineering skills~\citep{denny2023computing}. Coding tools such as GitHub's Copilot or OpenAI's Codex may lead novices to over-rely on auto-suggested solutions. This overreliance may diminish their engagement with computational thinking~\citep{Prather_2023, becker2022programming}, which is arguably the most important skill that is learned in any computer science course or program. 

Beyond this example, many studies underscore the significance of adapting course materials and assessments to promote critical thinking, encourage student collaboration, develop practical skills, enhance soft skills, and promote interdisciplinary learning, all with the aim of cultivating graduates equipped with a diverse range of competencies~\citep{nowrozy2024embracing, alier2024generative, Chaudhry, Cotton}. In particular, reinforcing our conclusions above, open-ended assessments are proposed to promote originality and creativity, potentially discouraging reliance on generative AI tools and fostering unique ideas and critical analysis~\citep{Cotton, Liu}. Ultimately, these studies suggest the greater risk of generative AI may be its potential to enable the unintentional circumvention of frameworks by which learners are taught the foundations to learn later skills, and that teaching and assessment should be adapted for this risk.

Finally, assuming that students will use and become acquainted with the capabilities of these technologies, we recommend that students should not only be educated on the technical and ethical challenges of generative AI systems, but also on the critical thinking required to successfully engage with such tools~\citep{wang2023examining}. One such measure could involve establishing committees for ethical oversight and adding classroom discussions on the use of AI tools. Such discussions would clarify what constitutes plagiarism and address potential ethical concerns, ensuring that students understand the importance of academic integrity and discern the boundaries between legitimate assistance and academic misconduct~\citep{Finnie-Ansley, Cotton,alier2024generative, Chaudhry, Liu,denny2023computing}.

\section{Limitations}
While our study offers preliminary insights into the vulnerability of degree programs to student use of AI assistants for assessments, we acknowledge several limitations in our study. 

First, our study excluded any multimodal inputs, such as questions containing diagrams, figures, or graphs, which were omitted from our dataset. Approximately 57\% of the initially collected data did not qualify for inclusion in the final dataset of 5,579 questions. Consequently, models were solely evaluated with text-only questions. This approach likely resulted in performance outcomes that are higher than what these models would attain when tested on question sets that include these other modalities, though we also note rapid growth in the multimodal capabilities of these models \citep{Yue2023MMMUAM}.

Simultaneously, our findings might underestimate the performance potential that students could attain through collaboration with these systems. Although we conducted a thorough examination of prompting strategies, our methods are limited by the fact that they (1) rely solely on published prompting strategies, (2) are generally non-interactive, and (3) are tailored for scalability across all questions to facilitate a comprehensive study. Students aiming to address individual questions could devote more time and devise more interactive, less standardized prompting strategies to collaboratively guide the models toward improved solutions. 

Finally, we acknowledge certain gaps between our evaluation of AI systems in this study, and how students are normally evaluated in these courses. First, our study standardizes system evaluation across all course assessments, removing course-specific assessment policies for questions. For example, certain courses, beyond not giving points for correct answers to multiple-choice questions, might also penalize incorrect answers more than leaving a question unanswered, while our study simply gives zero points for incorrect answers. Second, our dataset of questions for each course is not necessarily balanced according to a course's grading rubric. As an example, our dataset may contain a balanced mixture of questions from assignments and exams for a particular course, while the overall evaluation of a student in this same course would compute their grade as a 10\% mixture of assignments, and 90\% mixture of exam questions. Similarly, many courses at our institution also include lab or project components as part of their final grade. Since these parts of the assessment do not have a ``correct answer'' that can be easily marked, they are not included in our dataset. 

As we do not consider these course-specific assessment policies when computing the course pass rates of our tested AI assistants, these design decisions introduce a gap between our evaluation and the actual assessment rubrics by which students are graded in our institution's courses. Despite this divergence, however, we note that other institutions may implement course assessments and grading rubrics in different ways. As a result, while our study is not an exact simulation of our institution's diverse course assessment schemes, it remains a suitable testbed for providing insights into how course assessments are vulnerable to AI assistants, and how this vulnerability would extend to full university programs without mitigations.

\section{Materials and Methods}

\subsection{Prompting Strategies}
To generate answers to questions, we employ various prompting strategies requiring only familiarity with relevant literature and minimal adaptation. We selected eight distinct prompting strategies that we broadly categorize into three types: direct, rationalized, and reflective prompting. 
Under direct prompting, we use zero-shot, one-shot \citep{brown2020language}, and expert prompting \citep{xu2023expertprompting}, where models are directly prompted for an answer without encouraging any underlying rationale. For rationalized prompting, three strategies are implemented: zero-shot and four-shot chain-of-thought \citep{wei2023chainofthought}, and tree-of-thought \citep{yao2023tree} prompting. Here, language models are prompted to generate a rationale before providing the final answer. Lastly, reflective prompting includes self-critique \citep{wang2023selfcritique,madaan2023selfrefine} and metacognitive prompting \citep{wang2023metacognitive}, where models are asked to reflect on a previously provided answer and adjust their response according to this reflection. In our experiments, we noted that the prompting strategy substantially influences model performance, with at least one strategy consistently producing the correct answer in the majority of cases. A detailed description of all prompting strategies, along with prompts, is provided in Appendix \ref{ap:prompt}.
% {\color{red} SI 2}.

\vspace{-2mm}
\subsection{Evaluation}
Below, we outline the grading strategies used to evaluate the model's performance across two question types: multiple-choice (MCQ) and open-answer questions. For MCQ, grading is automated by comparing against the annotated answer. Answers receive a binary score of $0/1$ if only one correct option exists, or a proportional score based on the number of correct choices made for multi-answer questions (with no penalty for wrong choices). Appendix \ref{ap:grading_eval} provides more details for grading MCQs. For open-answer questions, we constructed an evaluation pipeline using GPT-4 as a grader~\citep{zheng2024judging}, which we describe below. For both types of results, we report error bars representing 95\% confidence intervals (Figures~\ref{fig:allignment}, \ref{fig:course-size}). These intervals were computed using the non-parametric bootstrap with 1000 resamples. We also tasked human experts with independently grading a subset of model responses to measure alignment between model and human grading, and establish a confidence level for model-based grading.

\vspace{2mm}
\noindent \textbf{Evaluating Open-Answer Questions.} A significant portion of the questions we extracted are open-answer questions, which are challenging to evaluate manually due to the sheer volume of responses (a total of 33,904 answers from 2,119 questions, answered by 2 models using 8 prompting strategies). As a result, we use a state-of-the-art LLM, GPT-4, as a grader. To automate the grading of open answers, we provide the model with the question, the correct solution from an answer sheet of the assessment, and the generated answer text, prompting it to assign a rating based on the quality of the response. We provide the model with a 4-point grading scale: \texttt{Correct, Mostly Correct, Mostly Incorrect, Incorrect}. The model is first tasked with assessing the accuracy and completeness of the answer before assigning the final grade. Although we do not use these interim accuracy and completeness scores, we manually observe that these stages enhance the quality of overall grading. The specific prompting strategy is detailed in Appendix \ref{ap:direct_grading}. As an answer was produced for each question using eight distinct prompting strategies, we obtained eight different answers and corresponding grades. To present a cohesive performance score for both GPT-4 and GPT-3.5 for a given question, we employ two aggregation methods: (1) the \textit{maximum} approach, which selects the answer with the highest grade for each question as a representation of model performance, and (2) the \textit{majority} approach, which considers the grade that appears most frequently among the eight prompting strategies. As an example, for a hypothetical question whose generated answers for the eight prompting strategies received 2 \texttt{Correct}, 2 \texttt{Mostly Correct} and 4 \texttt{Mostly Incorrect} grades, the \textit{maximum} grade would be \texttt{Correct} and the \textit{majority} grade would be \texttt{Mostly Incorrect}.

\begin{figure}
    \centering
    \includegraphics[width=\linewidth]{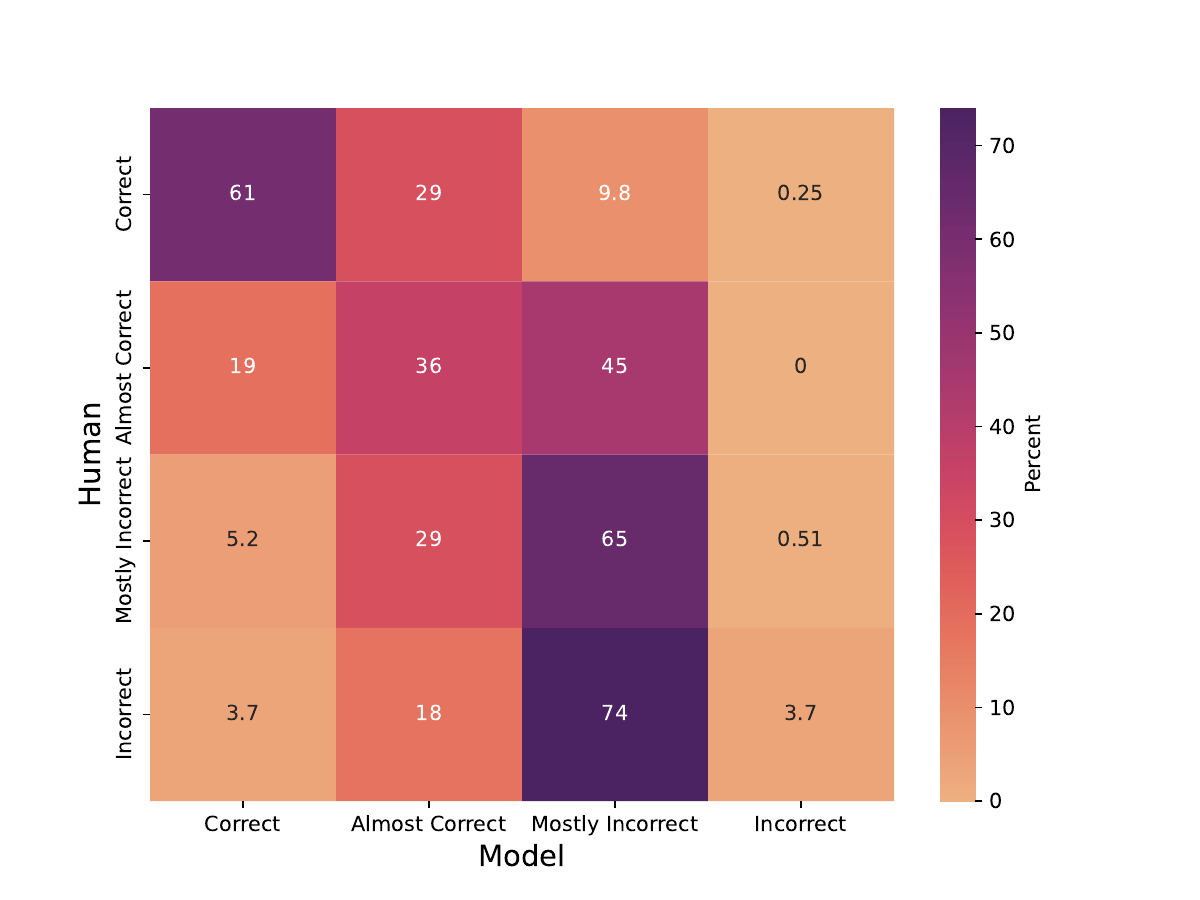}
    \caption{\textbf{Comparison of Human and GPT-4 grading.} Average model and human performance for 933 questions and responses from GPT-4 and GPT-3.5 generated with the metacognitive prompting method.}
    \label{fig:grading}
\end{figure}

%correlation of the model's final grades with humans graders.

\vspace{2mm}
\noindent \textbf{Human Grading.} To assess how well model grading aligned with human grading on open-answer questions, we enlisted 28 expert annotators from the teaching faculty of 11 courses to evaluate 933 questions. The courses chosen for expert grading are listed in Appendix \ref{ap:human_grading}. Specifically, we requested graders to assign scores to open-ended responses generated by GPT-4 and GPT-3.5. Human-graded responses for both models were generated using two prompting strategies: zero-shot chain-of-thought prompting \citep{wei2023chainofthought} (a simple prompting method at the disposal of any student) and metacognitive prompting \citep{wang2023metacognitive} (one of the most effective strategies across all courses). We anonymized the outputs to prevent graders from knowing which model and prompting strategy they were evaluating. To maintain consistency, we instructed graders to use the same grading scale as GPT-4 direct grading. The number of graders per course varied from 1 to 10, and a total of 3732 answers were evaluated. On average, graders spent approximately 5 minutes assessing each answer.

Figure \ref{fig:grading} indicates a general alignment between human graders and GPT-4 when categorizing answers into a simplified correct/incorrect quadrant. Out of the examples identified as \texttt{Correct} by graders, the model assigned the same grade to 61\% of them. Similarly, for examples graded as \texttt{Almost Correct} by graders, the model's grade matched in 36\% of cases. Additionally, in instances where graders labeled examples as \texttt{Mostly Incorrect}, the model's grade aligned with the grader's assessment 65\% of the time. However, we note certain discrepancies. For instance, GPT-4 tends to avoid explicitly labeling solutions as \texttt{Incorrect}, instead opting for \texttt{Mostly Incorrect} (i.e., in 74\% of cases that humans annotated a solution as \texttt{Incorrect}, the model identified it as \texttt{Mostly Incorrect}), potentially due to the practice of aligning models for harmlessness \citep{bai_constitutional_2022}. %Additionally, the model often categorizes solutions as \texttt{Mostly Incorrect} that humans deemed \texttt{Almost Correct} (i.e., In 45\% of cases that humans annotated a solution as \texttt{Almost Correct}, the model identified it as \texttt{Mostly Incorrect}). 
We find a few instances where the model rates an answer as \texttt{Correct} while humans assign a lower score. 

Interestingly, upon comparing average grades assigned by human graders and GPT-4 across 11 courses, we find a difference in average grade of only 2.75\%. However, we observe variations between courses, with an average course grade deviation of 8.5\% (and the largest deviation for a course being 26\%) between human and model graders. Finally, we also note the performance correlation between MCQ and open-answer questions in Figure \ref{fig:course-passing-rate}, providing a comparison point for the rationality of our model-based open-answer grading results. While scores for open-answer questions are typically lower than MCQ, the patterns exhibited by both curves are similar across both models. Overall, we note that the grades provided by humans and models are moderately correlated and that the summary statistics across courses tend to have a high correlation. Further details can be found in Appendix \ref{ap:grading_eval}. 

\begin{acks}
AB gratefully acknowledges the support of the Swiss National Science Foundation (No. 215390), Innosuisse (PFFS-21-29), the EPFL Science Seed Fund, the EPFL Center for Imaging, Sony Group Corporation, and the Allen Institute for AI. PS acknowledges support from the NCCR Catalysis (grant number 180544), a National Centre of Competence in Research funded by the Swiss National Science Foundation. TK is partially funded by the Swiss State Secretariat for Education, Research, and Innovation (No. 591711).
\end{acks}

\bibliographystyle{ACM-Reference-Format}
\bibliography{custom}

\appendix

\section{Dataset Collection}

\textbf{Data Statement} Our data collection was approved by an institutional review board. Data was voluntarily submitted by members of the Data Consortium and no materials were used without the permission of the data owner. 

\vspace{2mm}
\noindent \textbf{Data Preprocessing} To preprocess our data, we collect assessments from participating faculty, extract questions and answers from these assessments, and standardize them into a uniform format. After compiling an initial question bank from the raw data, we filter unsuitable data points by (1) removing questions that lack the question body or solution, (2) eliminating duplicate questions, and (3) removing questions that require information that cannot be parsed by LLMs in a textual format (e.g., diagrams, images, plots). In cases where a joint context is provided for multiple questions, we augment each question individually with this context.

\section{Prompting Strategies}\label{ap:prompt}
Our study's goal is to identify the vulnerability of educational assessments to AI systems. As a result, we select prompting strategies that simulate realistic student use and assess prompting strategies that can be used with minimum effort, requiring only knowledge of the relevant literature and minimal adaptation. We exclude strategies involving training models. Our assessment encompasses three primary categories of prompting strategies: \textit{direct prompting}, wherein the model is directly prompted to provide an answer; \textit{rationalized prompting}, which encourages the model to first verbalize reasoning steps before providing a response; and \textit{reflective prompting}, which prompts the model to reflect on a previously generated response before finalizing an answer. Each prompt is tailored for three scenarios: (1) MCQs with a single correct answer, (2) MCQs with multiple correct answers, and (3) open-answer questions. 
Below, we outline the strategies used to prompt models to answer questions:

\subsection{Direct Prompting}
We explore three strategies for direct prompting: zero-shot, one-shot, and expert prompting. All these strategies ask the LLM directly for an answer without encouraging any particular strategy or rationale to arrive at the answer.

\paragraph{Zero-shot Prompting.}
We ask the model to solve questions without any demonstrations or system role prompts. The instructions vary depending on the type of question: open-answer or multiple-choice (MCQ). Additionally, we differentiate between multiple-choice cases where a single answer is correct and cases where multiple correct answers can be selected.

\begin{prompt} % zeroshot + few-shot

\textbf{MCQ single answer}: You are given a question followed by the possible answers. Only one answer is correct. Output the correct answer.

\textbf{MCQ multi answer}: You are given a question followed by the possible answers. The question can have multiple correct choices. Output all the correct answers.

\textbf{Open answer}: Solve the following question:
\end{prompt}

\noindent For MCQs, each answer option is associated with a letter, and the model is expected to provide the letter corresponding to its choice.

\paragraph{One-shot Prompting \citep{brown2020language}.} In this prompting strategy, we instruct the model to solve questions based on a provided example as a demonstration, without any additional system role prompt. Each question is paired with a demonstration that is the most similar to the question being addressed. Specifically, we use the ``all-roberta-large-v1'' model\footnote{We use the sentence-transformers \citep{reimers-2019-sentence-bert} implementation of this model, available at \url{https://huggingface.co/sentence-transformers/all-roberta-large-v1}.} \citep{liu2019roberta} to embed all questions as vectors, and then retrieve the most similar question vector based on cosine similarity to the prompt question vector. We append the corresponding demonstrative example to this retrieved vector to the prompt. The prompt instructions remain the same as for zero-shot prompting. The demonstration is provided to the model in a multi-message setting, mimicking an actual conversation between the user and the assistant.

\paragraph{Expert Prompting \citep{xu2023expertprompting}.}
In expert prompting, we use the LLM to to simulate the responses of three experts in the field. The model generates answers as if written by these experts, and then we combine their responses using collaborative decision-making, typically through majority voting. This process is represented by using a generic expert defined as the system role, such as ``\textit{You are a professor of Machine Learning}'' for questions from, e.g., Machine Learning\footnote{Whenever a prompting strategy makes use of a course name, it is employing the course name rather than the course code (e.g., ``Machine learning for physicists'' rather than course code like ``PHYS 444'').} and prompting the model to give us the names of three experts in the field capable of solving the given question using the prompt:

\begin{prompt} % Expert Prompting
\textbf{System:} You are an expert in \{\textit{course name}\}.

% \textbf{Step 1:} 
Give an educated guess of the three experts most capable of solving the following question. Only output the name of these three experts as a json format with key as number and value as a name, without any explanation.
\end{prompt}

\noindent Following this, the model adopts the personas of the named experts as its system role to produce an answer, employing the same prompt as used in the zero-shot and one-shot strategies. The answers generated by these personas are then aggregated using a majority voting approach.

\subsection{Rationalized Prompting}
We explore three strategies for eliciting reasoned answers: zero-shot and four-shot chain-of-thought, and tree-of-thought prompting. Each strategy involves prompting the LLM to generate a rationale before providing a final answer.
\paragraph{Chain-of-Thought Prompting~(CoT) \citep{wei2023chainofthought}.}
In chain-of-thought prompting, we guide the model to generate a sequence of reasoning steps before providing an answer. This approach typically results in more coherent, structured, and accurate responses, as it requires the model to present arguments before delivering the final answer. This behavior is often initiated by an instruction such as ``Let’s think step by step.'' For better performance, the model may be given demonstrations that illustrate how to break down a question into multiple reasoning steps. However, manually generating these demonstrations for each course is time-consuming. Therefore, we automatically generate multiple example rationales using GPT-4 for questions from each course. Domain experts then manually select the best chain-of-thought reasoning trace for each question and correct or improve it if necessary.

\noindent We experiment with two settings: zero-shot (no demonstration) and few-shot (4 demonstrations). For the latter, for each question, we sample 4 demonstrations of the same course cluster (same topic) and the same question type (MCQ or open-answer), ensuring that these demonstrations were different from the question being asked to the model.
Sometimes, the total length of the 4 demonstrations exceeds the model's maximum context length. In such cases, we reduced the number of demonstrations to fit within the context limit. Additionally, we provided a system prompt that included the course topic as an extra hint for the model.
For each question type, the selected prompts are the following:
\begin{prompt} % CoT
\textbf{System:} You are an expert in \{\textit{course name}\}.

\textbf{MCQ single answer}: You are given a question followed by the possible answers. Only one answer is correct. Give a step-by-step reasoning, and then output the correct answer.

\textbf{MCQ multi answer}: You are given a question followed by the possible answers. The question can have multiple correct choices. Give a step-by-step reasoning, and then output all the correct answers.

\textbf{Open answer}: Solve the following question, by first giving the step-by-step reasoning and then outputting the answer:
\end{prompt}

\noindent The demonstration pairs, which include the question and its reasoning explanation, are provided to the OpenAI API using a multi-message setting similar to the few-shot strategy.

\paragraph{Tree-of-Thought Prompting \citep{yao2023tree}.}
While chain-of-thought prompting has led to performance improvements in many NLP tasks, it is sensitive to incorrect reasoning steps, as there is no mechanism to assess and fix a reasoning error after it has been made. Tree-of-thought prompting extends chain-of-thought by having the model emulate three subject experts. Each of them must generate a reasoning path and critique the other expert’s proposed paths. Then, the model is instructed to simulate a discussion between experts until they reach an agreement and provide a final answer. We use the following prompt to implement Tree-of-Thought:

\begin{prompt} % Tree-of-Thought
\textbf{System:} You are an expert in \{\textit{course name}\}.

Imagine three different experts answering this \{\textit{question type}\} question.
They will brainstorm the answer step by step reasoning carefully and taking all facts into consideration.~\\
All experts will write down one step of their thinking and then share it with the group.
They will each critique their response, and all the responses of others. They will check their answer based on science. Then all experts will go on to the next step and write down this step of their thinking. They will keep going through steps until they reach their conclusion taking into account the thoughts of the other experts. If at any time they realize that there is a flaw in their logic, they will backtrack to where that flaw occurred. If any expert realizes they're wrong at any point then they acknowledge this and start another tree of thought. Each expert will assign a likelihood of their current assertion being correct. Continue until the experts agree on the single most likely answer.
\end{prompt}

%The last part of the prompt differs slightly depending on the question type, similar to the previous strategies.

\subsection{Reflective Prompting}
We explore two strategies for reflective prompting: self-critique and metacognitive prompting. Both strategies involve the model reflecting on an answer it previously provided. Based on this reflection, the model then generates a final, improved answer.

\paragraph{Self-Reflect Prompting \citep{wang2023selfcritique,madaan2023selfrefine}.}

This strategy is performed on in conjunction to CoT to refine the reasoning traces generated by the model. Focusing on MCQ questions, first, we provide the model with a question and its zero-shot CoT response. Then, we prompt the model to revise its reasoning and produce a refined answer. Notably, this refinement process is carried out without any demonstrations.

\begin{prompt} % self-critique
\{\textit{CoT prompt and model output}\}

\textbf{MCQ single answer}: Please consider that there is a single correct choice. Is the provided reasoning accurate? If there isn't any inaccuracy, please output ``Reasoning is fine.'' Otherwise, please revise your reasoning and then choose the single correct choice.

\textbf{MCQ multi answer}: Please consider that multiple choices can be correct. Is the provided reasoning accurate? If there isn't any inaccuracy, please output ``Reasoning is fine.'' Otherwise, please revise your reasoning and then and then output all the correct choices.

\textbf{Open answer}: Assume you got the above answer from a student and you're looking for inaccuracies in either the reasoning or the final response. Try to refine any inaccuracy and answer the question from scratch. Please don't mention in your answer that you're refining a previous answer and write a new answer from scratch.
Answer: 
\end{prompt}

\paragraph{Metacognitive Prompting \citep{wang2023metacognitive}.}
Motivated by the concept of meta-cognition, this prompt is designed to emulate the human process of introspection and regulation of thinking. To achieve this, the language model is tasked with following a specific procedure akin to human cognitive processes. This involves sequentially: (1) deeply understanding the problem, akin to human comprehension; (2) identifying relevant concepts and formulating a preliminary answer; (3) evaluating and adjusting this preliminary answer if needed; and (4) confirming the final response and presenting it in a specified format.

\begin{prompt} % Metacognitive
You have to answer the following \{\textit{question type}\} question.~\\
\{\textit{Question text}\}~\\
As you perform this task, follow these steps:~\\
1. Clarify your understanding of the question.~\\
2. Make a preliminary identification of relevant concepts and strategies necessary to answer this question, and propose an answer.~\\
3. Critically assess your preliminary analysis. If you are unsure about its correctness, try to reassess the problem.~\\
4. Confirm your final answer and explain the reasoning behind your choice.
\end{prompt}

%\noindent Similar to tree-of-thought prompting, the last part of the prompt differs slightly depending on the question type.

\section{Evaluation}\label{ap:grading_eval}
In this section we describe the methods used for grading MCQs and open answer questions with GPT-4.

\subsection{Multiple Choice Scoring}

Regardless of the prompting strategy used for MCQs, the model is provided with the list of answer choices, each associated with a letter, and is asked to generate the letter(s) corresponding to the correct answer(s). Therefore, grading MCQs involves extracting the letter(s) indicated in the model's response and comparing them with the correct answer(s) (i.e., ground truth).

This process is straightforward for direct prompting strategies, but more challenging for strategies involving reasoning, such as chain-of-thought, where the model's response may include long explanations that discuss incorrect answers. To ensure consistency in answer extraction across different types of responses, we use an LLM (GPT-3.5) with the following prompt to extract the model's final answer:

\begin{prompt}
\textit{\{Question prompt\}}~\\
\textit{\{Model output\}}~\\
If the above answer does not provide an option, or gives an answer which is not in the options list, you should give the following:~\\
\{``selection'': [None]\}~\\
Otherwise, please return the answer in a dictionary format, with the key being ``selection'', and the value is a list that contains the index of letters of all the correct choices, with A being 0, B being 1, and so on:
  \end{prompt}

\subsection{Open Answer Direct Grading} \label{ap:direct_grading} 
For open-answer grading, we compare the performance of GPT-4 as a grader against human graders from the teaching staff of the courses from which the questions originated.

% \begin{figure}
%     \centering
%     \includegraphics[width=1\linewidth]{figures/new_figures/Human_Model.pdf}
%     \caption{\textbf{Comparison of human and GPT-4 grading.} Average model and human performance for a subset of 933 questions and answers generated with the metacognitive prompting strategy.}
%     \label{fig:grading}
% \end{figure}

\paragraph{Grading Open Answer Questions.}
To automatically grade the quality of open answers, the GPT-4 grader model is given the question, the gold solution (extracted from the course materials), and the text of the generated answer, and prompted to assign a rating to the generated answer based on its quality. 
Rather than asking the model for a single \textit{correct} or \textit{incorrect} label, we provide the model with a 4-point grading scale, ranging from \textit{correct}, \textit{almost correct}, \textit{mostly incorrect}, to \textit{wrong answer}. The full prompt is presented below:

\begin{prompt}
\textbf{System prompt}: 
You are a teacher of \textit{\{course name}\}. You must grade exam questions.\\\\
\textbf{User Prompt}:
You must rigorously grade an exam question. Please be strict and precise in your assessment, providing reasoning for your assigned grade. Here's the process I'd like you to follow:\\
Carefully read and understand the question.\\
Thoroughly compare the student's answer with the correct golden answer.\\
Evaluate the student's response based on its accuracy and completeness.\\
Deduce a final grade by considering whether the answer is ``wrong answer'', ``mostly incorrect'', ``almost correct'', ``correct'', along with a clear explanation for your decision. \\
Question: \textit{\{question\}}\\
Gold Answer: \textit{\{gold answer\}}\\
Student Answer: \textit{\{model output\}}\\
format your answer in the following json format, providing a clear and detailed evaluation for each of the two criteria (accuracy and completeness) and finally providing the grade. in the field of grade only write the final grade from the given grading options:\\
\{``accuracy'': ,\\
 ``completeness'': ,\\
 ``grade'':\\
 \}
\end{prompt}

\subsection{Comparing GPT-4 and Human Grading}\label{ap:human_grading}

To better understand GPT4's capabilities as a grader, we compare its grading performance against the human grading scores, using two metrics: \textit{Average Grade} and \textit{Grade Agreement}. We recruited 28 graders from 11 of the courses in our dataset and tasked them with providing a general assessment of the quality of 933 responses provided by GPT-3.5 and GPT-4. Similar to GPT-4 as a grader, human graders are asked to use the same 4-point scale to grade model outputs. Given the cost of performing this annotation, we only task graders to mark responses from two prompting strategies, Zero-shot CoT \citep{wei2023chainofthought} and Metacognitive prompting \citep{wang2023metacognitive}.

\begin{table*}[t]
\resizebox{\linewidth}{!}{
\centering
\begin{tabular}{ l r  llll  llll}
\toprule
   \multirow{3}{*}{\textbf{Course Name}} & \textbf{Prompting Strategy:} & \multicolumn{4}{c}{\textbf{Zero-Shot CoT}}  & \multicolumn{4}{c}{\textbf{Metacognitive Prompting}}  \\%\toprule
   & \textbf{Model:} & \multicolumn{2}{c}{GPT-4 Responses} & \multicolumn{2}{c}{GPT-3.5 Responses} &  \multicolumn{2}{c}{GPT-4 Responses} & \multicolumn{2}{c}{GPT-3.5 Responses}  \\
   & \textbf{Grader:}    & \multicolumn{1}{c}{Human} & \multicolumn{1}{c}{GPT-4} & \multicolumn{1}{c}{Human} & \multicolumn{1}{c}{GPT-4} & \multicolumn{1}{c}{Human} & \multicolumn{1}{c}{GPT-4} & \multicolumn{1}{c}{Human} & \multicolumn{1}{c}{GPT-4} \\  
\toprule
\multicolumn{2}{l}{Statistical Physics} &  $48.9  \pm 10.1 $ & $  53.7  \pm 6.9$ & $43.9  \pm 11.4 $ & $  38.6  \pm 4.4$ & $36.4  \pm 10.1 $ & $  45.5  \pm 6.3$ & $37.6  \pm 10.6 $ & $  39.9  \pm 4.4$ \\
%0.48 $ & $  0.53  &  0.43 $ & $  0.38  & 0.36 $ & $  0.45  &  0.37 $ & $  0.39 \\
\multicolumn{2}{l}{Concurrency \& Parallel Processing}       &$62.3  \pm 14.5 $ & $  68.4  \pm 9.5$ & $62.3  \pm 14.5 $ & $  61.0  \pm 8.1$ & $72.8  \pm 14.5 $ & $  68.4  \pm 10.5$ & $56.1  \pm 15.6 $ & $  53.8  \pm 9.4$ \\
%0.62 $ & $  0.68  &  0.62 $ & $  0.61  & 0.72 $ & $  0.68  &  0.56 $ & $  0.53 \\      
\multicolumn{2}{l}{Advanced Computer Architecture} &   $50.4  \pm 10.5 $ & $  74.9  \pm 6.2$ & $44.9 \pm 11.1 $ & $  68.8  \pm 6.9$ & $60.3  \pm 11.0 $ & $ 73.1  \pm 6.2$ & $42.3  \pm 10.5 $ & $  62.6  \pm 6.2$ \\           
%0.50 $ & $  0.74  &  0.44 $ & $  0.68  & 0.60 $ & $  0.73  &  0.44 $ & $  0.62 \\
\multicolumn{2}{l}{Software Engineering} &  $62.2  \pm 9.0 $ & $  85.9  \pm 4.8$ & $47.9  \pm 9.5 $ & $  72.9  \pm 5.3$ & $66.1  \pm 9.9 $ & $  84.1  \pm 5.8$ & $49.8  \pm 10.5 $ & $  73.0  \pm 5.7$ \\
%0.62 $ & $  0.85  &  0.47 $ & $  0.72  & 0.66 $ & $  0.84  &  0.49 $ & $  0.73 \\
\multicolumn{2}{l}{Mathematics of Data} &  $52.1  \pm 13.5 $ & $  65.4  \pm 8.1$ & $50.2  \pm 13.5 $ & $  56.4  \pm 8.1$ & $94.5  \pm 5.5 $ & $  68.9 \pm 7.3$ & $76.5  \pm 12.7$ & $  56.4  \pm 9.0$ \\
%0.52 $ & $  0.65  &  0.50 $ & $  0.56  & 0.94 $ & $  0.68  &  0.76 $ & $  0.56 \\
\multicolumn{2}{l}{ML for Physicists} &  $80.8  \pm 7.2 $ & $  76.7  \pm 5.2$ & $71.7  \pm 7.9 $ & $  69.1  \pm 6.3$ & $80.4  \pm 6.8 $ & $  73.9  \pm 4.8$ & $74.4  \pm 7.6 $ & $  69.5  \pm 5.2$ \\
%0.80 $ & $  0.76  &  0.71 $ & $  0.69  & 0.80 $ & $  0.73  &  0.74 $ & $  0.69 \\
\multicolumn{2}{l}{Semiconductor Properties} &  $74.1  \pm 15.3 $ & $  78.6  \pm 10.7$ & $66.4  \pm 12.2 $ & $  78.5  \pm 12.2$ & $63.4  \pm 16.6 $ & $  66.3  \pm 10.6$ & $55.8  \pm 15.2 $ & $  64.8  \pm 10.7$ \\
%0.74 $ & $  0.78  &  0.66 $ & $  0.78  & 0.63 $ & $  0.66  &  0.55 $ & $  0.64 \\
\multicolumn{2}{l}{Applied Data Analysis} & $74.8  \pm 13.1 $ & $  72.3  \pm 10.8$ & $58.1  \pm 13.1 $ & $  57.9  \pm 8.3$ & $73.6  \pm 13.1 $ & $  65.1  \pm 9.5$ & $65.2  \pm 13.1 $ & $  55.6  \pm 9.6$ \\
%0.74 $ & $  0.72  &  0.58 $ & $  0.57  & 0.73 $ & $  0.65  &  0.65 $ & $  0.55 \\
\multicolumn{2}{l}{Advanced General Chemistry} &  $78.9  \pm 8.4 $ & $  80.7  \pm 6.6$ & $58.8  \pm 10.8 $ & $  64.0  \pm 7.8$ & $80.8  \pm 7.8$ & $  81.4  \pm 7.8$ & $62.3  \pm 10.1 $ & $  62.3  \pm 7.8$ \\
%0.78 $ & $  0.80  &  0.58 $ & $  0.63  & 0.80 $ & $  0.81  &  0.62 $ & $  0.62 \\
\multicolumn{2}{l}{Information \& Communication} & $76.6  \pm 4.7 $ & $  74.3  \pm 3.9$ & $57.6  \pm 5.1 $ & $  56.2  \pm 3.9$ & $74.3  \pm 4.1 $ & $  70.9  \pm 3.9$ & $60.3  \pm 5.1 $ & $  59.5  \pm 3.9$ \\ 
%0.76 $ & $  0.74	&  0.50 $ & $  0.56  &  0.74 $ & $  0.70  &  0.60 $ & $  0.59 \\
\multicolumn{2}{l}{Analysis I} &   $37.2  \pm 4.7 $ & $  48.7 \pm 3.2$ & $28.8  \pm 4.1 $ & $  38.4  \pm 2.6$ & $48.6  \pm 4.1 $ & $  47.1  \pm 3.1$ & $42.9  \pm 4.3 $ & $  44.5  \pm 2.8$ \\
%0.37 $ & $  0.48	&  0.30 $ & $  0.40	&  0.50 $ & $  0.49  &  0.41 $ & $  0.43 \\
\midrule
\multicolumn{2}{l}{Average} &     $63.5 \pm 9.1 $ & $  70.9 \pm 6.9$  &  $57.7 \pm 10.3 $ & $  60.2 \pm 6.7$  & $68.3 \pm 9.4$ & $  67.7 \pm 6.9$  &  $56.7 \pm 10.5$ & $  58.4 \pm 6.8$\\
\bottomrule
\end{tabular}
}\caption{\textbf{Comparison between human graders and the GPT-4 model across multiple university courses.} { The average grades provided by human graders and the GPT-4 model for open-answer questions. Results are presented for two prompting strategies (Zero-shot CoT and Metacognitive) and each student model (GPT-3.5 and GPT-4). Each performance is reported with a 95\% confidence interval.}}\label{table:grades}
\end{table*}

\paragraph{Average Grade.} 
To evaluate the similarity between grades given by humans and GPT-4 for each course, we first compare the average grades they provide to responses to questions in each course of our dataset. To quantify the grades given by the model and humans, we map grade ratings to a discrete range between 0 and 1: \{\textit{correct}: 1.0, \textit{almost correct}: 0.66, \textit{mostly incorrect}: 0.33, \textit{wrong answer}: 0.0\}.

Table~\ref{table:grades} shows the average grades provided by both human graders and GPT-4 to question responses generated by both GPT-4 and GPT-3.5. Two prompting strategies were used: zero-shot CoT and metacognitive prompting. On average, for most of the courses, the model tends to give higher grades compared to human graders, particularly for the zero-shot CoT prompting strategy. Some courses show a significant disparity, with GPT-4 giving much higher grades than humans (e.g., \textit{Advanced Computer Architecture} and \textit{Software Engineering}). More rarely, the human graders consistently give higher scores for courses such as \textit{Machine Learning for Physicists} or, to a lesser extent, \textit{Applied Data Analysis}. We also observe variations between the two prompting strategies: for \textit{Mathematics of Data}, humans give higher grades than the model for metacognitive prompting, while the model gives the highest grades for zero-shot CoT. Overall, both GPT-4 and human graders tend to give higher grades to GPT-4 answers than GPT-3.5. Despite these differences, these results also indicate that humans and GPT-4 have a similar grading distribution for model responses (particularly for responses to the metacognitive prompting strategy).

\begin{table}[h]
\centering
\resizebox{\linewidth}{!}{
\begin{tabular}{ l  cc  cc }
\toprule
 \multirow{3}{*}{\textbf{Course Name}} & \multicolumn{4}{c}{\textbf{Pairwise Agreement (\%)}} \\
    & \multicolumn{2}{c}{Zero-Shot CoT}  & \multicolumn{2}{c}{Metacognitive} \\%\toprule
   & GPT-4 & GPT-3.5 &  GPT-4 & GPT-3.5 \\
\midrule
Statistical Physics of Computation &           26.4  &  20.8  & 22.6  &  18.9 \\
Concurrency and Parallelism &          37.5  &  40.6  &  40.6  &  18.7\\
Advanced Computer Architecture &               35.2  &  24.1  & 31.5  &  22.2 \\
Software Engineering &                         50.0  &  37.1  & 50.0  &  30.0 \\
Mathematics of Data &                          29.7  &  18.9  & 32.4  &  24.3 \\
ML for Physicists &                            54.8  &  51.2  & 52.4  &  42.9 \\
Semiconductor Properties &                     59.1  &  31.8  & 31.8  &  50.0 \\
Applied Data Analysis &                        35.7  &  42.9  & 35.7  &  32.1 \\
Advanced General Chemistry &                   62.5  &  41.1  & 71.4  &  48.2 \\
Information and Communication &  59.1  &  48.0  & 61.8  &  53.8 \\
Analysis I &                                   42.6  &  37.6  & 44.6  &  37.6 \\
\midrule
Average &                                      44.8  &  35.8  & 43.2  &  34.4 \\
\bottomrule
\end{tabular}
}
\caption{\textbf{Pairwise agreement (\%)} between grades provided by human graders and the GPT-4 model as a grader.}\label{table:grade_agreement}
\end{table}

\begin{figure*}[ht]
    \centering
    \begin{subfigure}[b]{0.45\linewidth}  % Adjust the width to fit side by side
        \centering
        \includegraphics[width=\textwidth]{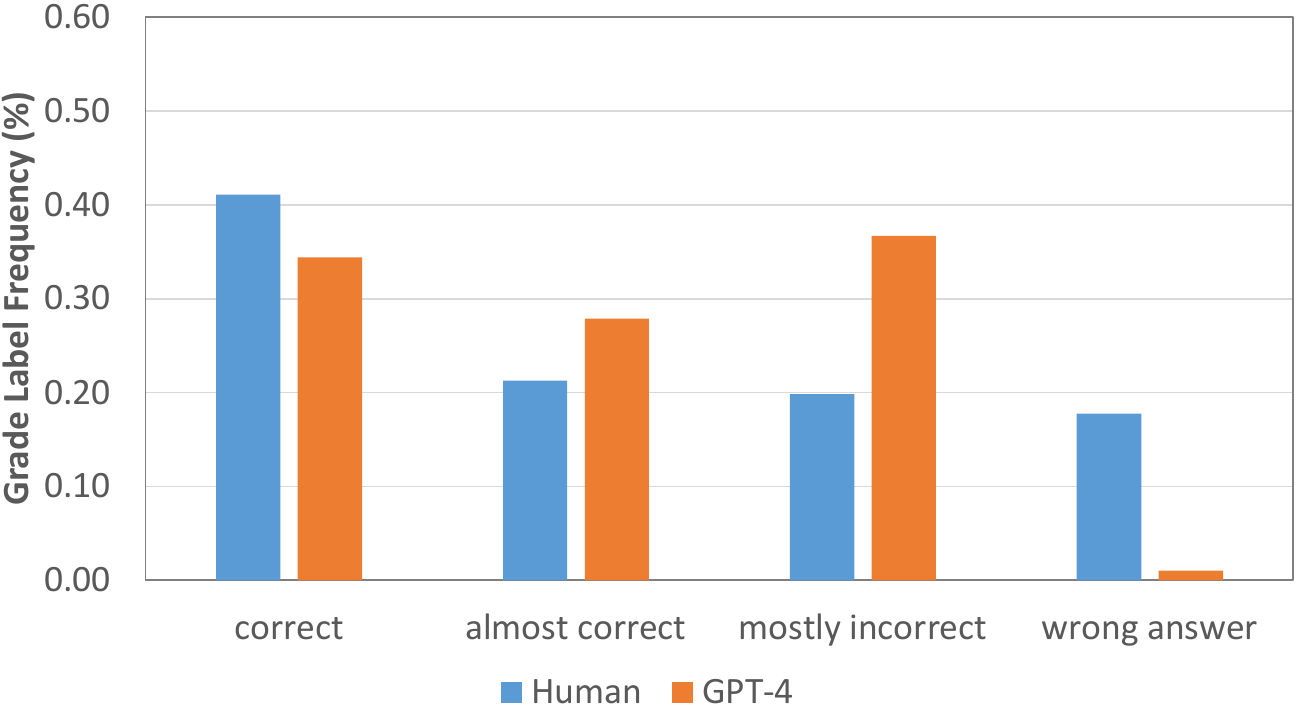}
        \caption{GPT-4 Student}
        \label{fig:GPT-4_label_dist}
    \end{subfigure}
    \hfill
    \begin{subfigure}[b]{0.45\linewidth}  % Adjust the width to fit side by side
        \centering
        \includegraphics[width=\textwidth]{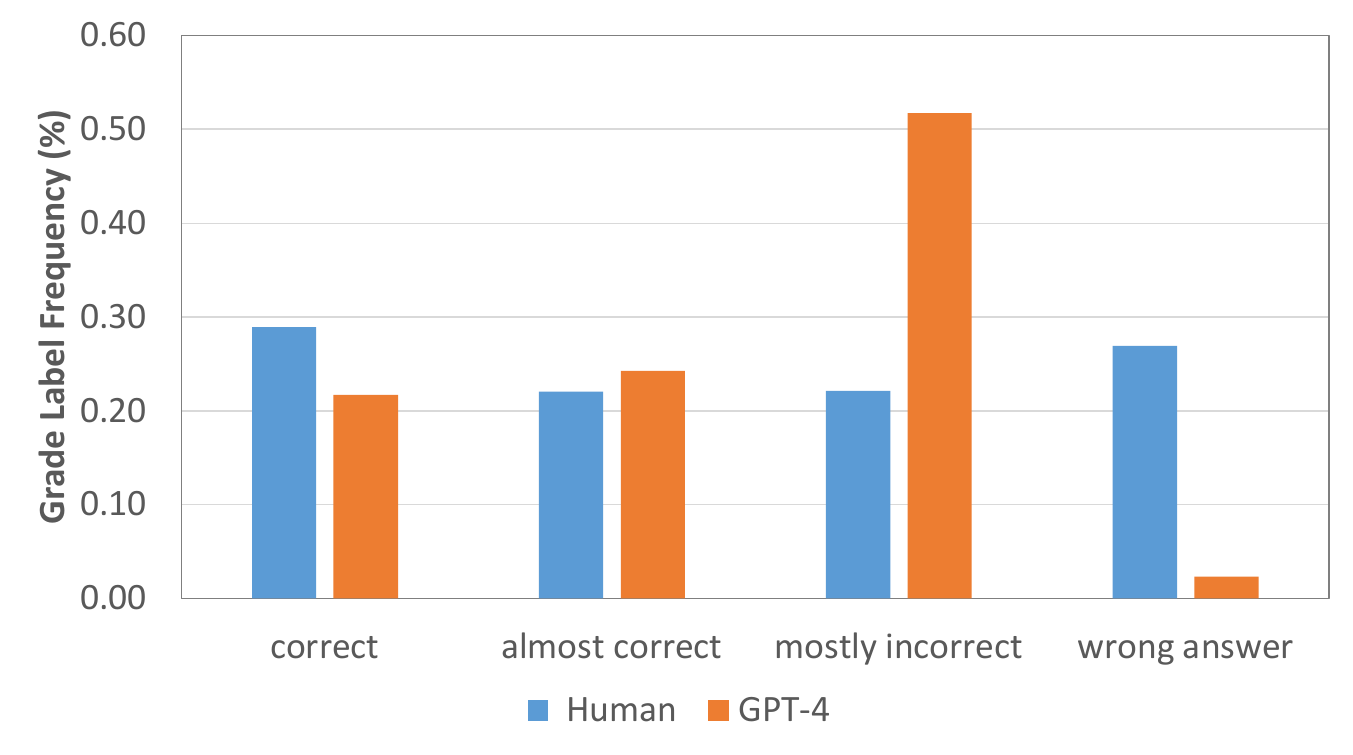}
        \caption{GPT-3.5 Student}
        \label{fig:chatgpt_label_dist}
    \end{subfigure}
    \caption{\textbf{Grade label distributions.} Distribution of grades assigned by our grader consortium (blue) and GPT-4 (orange) to the responses provided by GPT-4 and GPT-3.5.}
    \label{fig:label_distribution}
\end{figure*}

\paragraph{Agreement.} While investigating average grades provides an initial assessment of whether GPT-4's grading distribution generally matches that of humans, it does not give us a comprehensive understanding of the alignment between GPT-4 and the teaching staff's grading, so we now investigate the level of response-level grade agreement between the two. The agreement is defined as the percentage of question responses for which the model and human give the same grade. Table~\ref{table:grade_agreement} shows the average rate of grade agreement for each course. For each course, student model, and prompting method, we report the exact agreement between human graders and GPT-4 as a grader. The agreement between the model and the human varies for different courses, changing from  $\sim$18\% to $\sim$70\%, while the average agreement across all courses stays below 50\% for both metacognitive and zero-shot CoT. Figure~\ref{fig:label_distribution} shows the human and GPT-4 assigned grades distribution for both GPT-4 and GPT-3.5 as the student, including the two prompting strategies.
We observe that GPT-4 as a grader tends to grade model outputs using the labels \textit{almost correct} and \textit{mostly incorrect} far more often than human graders, while rarely identifying a response as \textit{wrong answer}. In contrast, human graders are more generous at identifying responses as \textit{correct}, but almost more willing to identify responses as \textit{wrong answer}.

\paragraph{Human Grader Remarks.} During the human grading process, we asked the 28 graders to record their impressions of model answers. Overall, there was a general agreement that the model's responses were satisfactory for straightforward questions, but less so for those requiring logical reasoning or analysis. In the latter cases, it was noted that the model sometimes produced lengthy responses that added contextually relevant information but failed to actually solve the problem. In many instances, graders likened this behavior to students attempting to gain points by including all potentially relevant information related to a question's keywords. This behavior could also be an artifact of the prompting approach, however, as we used metacognitive and chain-of-thought prompting strategies to generate the outputs provided to human graders. While these strategies have the best performance on MCQ, they also tend to produce longer answers to open-ended questions. 

Other issues identified include instances of factual inaccuracies (e.g., fabricated references) and contextual inaccuracies (e.g., using concepts unsuitable for the requested analysis). Finally, the model, at times, misunderstood the objective of the question (e.g., providing an implementation-specific answer when a student would instead interpret it as a design question). Regarding mathematical reasoning, apart from the previously mentioned limitations, the models struggled significantly with mathematical derivations requiring multiple steps, demonstrated a flawed understanding of imaginary numbers, and made errors in calculations.

% \begin{figure}[h]
%     \footnotesize 
%     \centering
%     \subfloat[GPT-4 Student]{
%         \label{fig:GPT-4_label_dist}
%         {\includegraphics[width=0.5\columnwidth]{gpt-4_label_distribution.pdf}}
%     }
%     \subfloat[GPT-3.5 Student]{
%         \label{fig:chatgpt_label_dist}
%         {\includegraphics[width=0.5\columnwidth]{chatgpt_label_distribution.pdf}}
%     }
%     \caption{\textbf{Grade label distributions.} Distribution of grades assigned by our grader consortium (blue) and GPT-4 (orange) to the responses provided by GPT-4 and GPT-3.5.}
%     \label{fig:label_distribution}
% \end{figure}

\subsection{Automated Grading in Prior Work} A substantial body of research leverages Large Language Models (LLMs) for response evaluation. Traditionally, automated assessment has necessitated high-quality reference data obtained through human grading, which is both costly and time-intensive. Consequently, there has been considerable exploration into the potential of LLMs to serve as evaluators \citep{chiang2023large}. Recent research has found LLMs to be capable of generating quality feedback \citep{scheurer2022training, welleck2022generating, tandon-etal-2022-learning, saunders2022selfcritiquing, paul2023refiner, schick2022peer, chen2023teaching, madaan2023selfrefine}, a trend also reflected in investigations into LLM-based evaluation \citep{fu2023gptscore, kocmi2023large, wang2023chatgpt, liu2023geval, zheng2024judging}. 

Automated solutions for student grading have been explored in the field of learning science, as well, some of which now use LLMs \citep{hasanbeig2023allure}. Intelligent Tutoring Systems (ITS), such ALEKS \cite{aleks}, ASSISTments \cite{Heffernan2014}, Cognitive Tutor \cite{Ritter2015CarnegieLC}, and MATHia \cite{ritter2011research} are widely employed to automatically assess student performance in closed-ended questioning. These systems cater to several hundred thousand students annually \cite{Heffernan2014, aleven2006toward}. Meanwhile, Automated Essay Scoring (AES) platforms such as e-Rater \cite{Attali_Burstein_2006}, IntelliMetric \cite{rudner2006evaluation}, and Intelligent Essay Assessor \cite{foltz1999intelligent} have emerged as useful tools for evaluating numerous student essays and responses to open-ended questions each year \cite{shermis2013handbook, rudner2006evaluation, foltz1999intelligent, ramesh2022automated, beigman-klebanov-madnani-2020-automated}.

\section{OpenAI API Hyperparameters}
For both GPT-4 and GPT-3.5, we set {\ttfamily temperature=0.8} to increase diversity and encourage more creative responses.
To reduce repetitive samples and keep the quality of the generations high, we set {\ttfamily presence\_penalty=0.5} and  {\ttfamily frequency\_penalty=0.8}. These values are chosen based on a human evaluation of the fluency and quality of the responses given a set of questions. For the rest of the hyperparameters, we use their default values.

\section{Bloom's Taxonomy}\label{ap:bloom}
Bloom’s taxonomy~\citep{blooms_rating} is a framework for categorizing learning objectives and educational items into levels of complexity requiring different cognitive skills. The taxonomy consists of 6 levels, from basic knowledge recall to higher-order critical thinking. The lower levels (\textit{remember}, \textit{understand}, and \textit{apply}) focus on foundational cognitive tasks such as remembering facts and comprehending basic information. As learners progress to higher-level categories, they engage in more complex cognitive tasks. The upper levels (\textit{analyze}, \textit{evaluate}, and \textit{create}) emphasize critical thinking, problem-solving, and creativity.

Although Bloom's Taxonomy is widely accepted and used, educators often disagree about the precise definitions of each category. This discrepancy leads to varied interpretations and challenges in categorizing learning objectives and educational items into specific taxonomy levels. This is particularly true when moving between adjacent levels, such as \textit{understand} and \textit{apply}. For example, given the following MCQ: 
\begin{prompt}
\texttt{In which of the following cases does JOS acquire the big kernel lock?}\\
\texttt{Options:}\\
\texttt{A. Processor traps in user mode}\\
\texttt{B. Processor traps in kernel mode}\\
\texttt{C. Switching from kernel mode to user mode}\\
\texttt{D. Initialization of application processor} 
\end{prompt}

On the one hand, to solve this question correctly, it is required to recall specific knowledge (\textit{remember}) about the circumstances under which JOS acquires the big kernel lock from the lecture or other learning materials. However, it can also be classified as the \textit{understand} category, as some multiple-choice options act as distractors that test the depth of a student's comprehension of the topic. As a result, the taxonomy has limitations in addressing the complexities of modern learning environments, especially in blended learning where information access and processing diverge from the conventional classroom setting for which Bloom's Taxonomy was crafted. 

Despite these ambiguities, Bloom's taxonomy remains a leading categorization scheme of cognitive difficulty in education. In this work, we assign Bloom's taxonomy labels to various questions in our dataset to assess model performance across questions of varying cognitive difficulty. To assign Bloom's meta-labels to questions in our dataset, we tasked two experts in the learning sciences to label 207 randomly-selected English questions with one of the six Bloom categories. They achieved an inter-annotator agreement of 51\% on this task. Using a more forgiving \textit{fuzzy agreement} (which also indicates an agreement if the annotators select adjacent categories) yields an agreement score of 84\%. Results for performance stratified by Bloom's taxonomy label can be found in the main article. 

\section{Program Statistics}\label{ap:program_stat}
In our work, we study 9 programs from three program levels: Bachelor, Master, and Online. Table \ref{tab:program-stat2} shows the number of courses available per program.
\begin{table}[t]
    \centering
    \begin{tabular}{lccccc}
    \toprule
    \multirow{2}{*}{\textbf{Program}} & \multicolumn{3}{c}{\textbf{Number of Courses}}\\
    &\textbf{Required}& \textbf{Optional} & \textbf{Total}\\
    \midrule
    Engineering, 1st year BSc & 5/12 & - & 5\\
    Chemistry, 1st year BSc & 6/9  & - & 6\\
    Life Science, 1st year BSc & 7/11 & - & 7\\
    \midrule
    Physics, BSc & 8/26& 1& 9\\
    Computer Science, BSc & 10/21 & 2&12\\
    \midrule
    Computer Science, MSc& 5/10& 3&8\\
    Data Science, MSc & 3/9 & 7&10\\
    \midrule
    Physics, Online & -& -&11\\
    Life Sciences, Online & -&-&8\\
    \bottomrule
    \end{tabular}
    \vspace{2pt}
    \caption{\textbf{Program Statistics.} 
    {``Required'' shows the ratio of required courses present in our data over the total number of required courses per program. ``Optional'' shows the number of optional courses per program. ``Total'' shows their sum, that is, the total number of courses our dataset covers, per program.}}\label{tab:program-stat2}
\end{table}

\section{Additional Results}\label{ap:additional_results}

\subsection{Individual Course Performance}\label{ap:course}
Table \ref{tab:gpt4_openanswer} shows GPT-4 performance across all courses for open-answer questions. Table \ref{tab:gpt4_mcq} shows GPT-4 performance across all courses for MCQ type of questions. Table \ref{tab:gpt3_openanswer} show GPT-3.5 performance across all courses for open-answer type of questions. Table \ref{tab:gpt3_mcq} show GPT-3.5 performance across all courses for MCQ type of questions. 
As the exact course names are not important for this analysis, we anonymize course names when presenting results in these Tables. 

\begin{table*}
 \centering
 \fontsize{9.0pt}{\baselineskip}\selectfont % font size
 \renewcommand\tabcolsep{2.5pt} % column space
 \renewcommand\arraystretch{0.85} % row space
  \resizebox{\linewidth}{!}{

 \begin{tabular}{l crrrrrrrrrr}
 \toprule
  \multirow{2}{*}{\textbf{Course name}}& \multirow{2}{*}{\textbf{\# questions}}&\multirow{2}{*}{\textbf{Zero-Shot}} & \multirow{2}{*}{\textbf{One-Shot}} & \textbf{CoT} & \textbf{CoT} & {\textbf{Tree-of-}} & {\textbf{Meta-}} & \multirow{2}{*}{\textbf{Expert}} & \multirow{2}{*}{\textbf{Self-Reflect}} & \multirow{2}{*}{\textbf{Majority}}& \multirow{2}{*}{\textbf{Max}}\\
&&&& (Zero-Shot) & (Four-Shot) & \textbf{Thought}& \textbf{cognitive} & & \\
\midrule
$^*$Biology \#1 & 12 & 80.3 & 66.3 & 83.1 & 85.8 & 77.5 & 83.0 & 69.1 & 91.5 & 85.9 & 94.3 \\
\midrule
$^*$Chemistry \#1 & 74 & 69.9 & 69.9 & 79.9 & 80.0 & 75.4 & 80.9 & 75.4 & 82.7 & 80.9 & 94.0 \\
\midrule
$^*$Computer Science \#1 & 42 & 61.6 & 66.4 & 63.9 & 65.6 & 63.9 & 61.6 & 67.1 & 73.6 & 65.6 & 88.7 \\
Computer Science \#2 & 98 & 65.0 & 69.1 & 67.8 & 66.4 & 65.3 & 64.3 & 67.4 & 74.6 & 66.3 & 89.6 \\
$^*$Computer Science \#3 & 42 & 64.8 & 67.2 & 75.2 & 78.3 & 67.9 & 64.8 & 74.3 & 75.9 & 70.3 & 93.5 \\
$^*$Computer Science \#4 & 42 & 76.8 & 69.6 & 78.3 & 74.3 & 71.1 & 77.6 & 83.1 & 80.0 & 79.2 & 97.6 \\
$^*$Computer Science \#5 & 223 & 68.5 & 69.7 & 74.3 & 71.5 & 68.6 & 70.9 & 73.4 & 73.4 & 73.1 & 89.4 \\
$^*$Computer Science \#6 & 72 & 54.2 & 57.5 & 53.3 & 52.4 & 51.5 & 51.5 & 57.5 & 59.4 & 53.8 & 79.3 \\
$^*$Computer Science \#7 & 70 & 81.1 & 82.1 & 85.9 & 88.8 & 80.2 & 84.1 & 88.4 & 89.4 & 87.4 & 98.1 \\
Computer Science \#8 & 36 & 57.1 & 62.6 & 70.0 & 65.4 & 53.3 & 58.0 & 59.8 & 70.0 & 57.1 & 89.7 \\
$^*$Computer Science \#9 & 55 & 65.7 & 67.5 & 75.4 & 78.5 & 65.0 & 73.0 & 73.0 & 79.7 & 74.8 & 90.1 \\
$^*$Computer Science \#10 & 34 & 82.1 & 80.1 & 86.1 & 83.1 & 84.1 & 78.2 & 89.1 & 90.1 & 87.1 & 98.0 \\
\midrule
$^*$Data Science \#1 & 28 & 71.1 & 67.5 & 72.3 & 71.1 & 69.8 & 65.1 & 66.3 & 72.3 & 68.7 & 90.3 \\
Data Science \#2 & 36 & 73.7 & 68.1 & 65.3 & 66.3 & 71.0 & 69.1 & 72.8 & 71.0 & 70.9 & 91.5 \\
\midrule
$^*$Math \#1 & 103 & 46.9 & 45.9 & 48.2 & 39.5 & 54.3 & 49.1 & 51.8 & 51.4 & 47.2 & 74.4 \\
$^*$Math \#2 & 302 & 60.9 & 60.0 & 58.7 & 56.9 & 64.7 & 66.1 & 64.9 & 62.9 & 62.3 & 83.1 \\
\midrule
Online Life Sciences \#1 & 9 & 36.7 & 40.4 & 66.2 & 70.1 & 58.9 & 59.0 & 58.9 & 62.7 & 58.9 & 92.6 \\
Online Life Sciences \#2 & 1 & 100.0 & 100.0 & 100.0 & 100.0 & 100.0 & 100.0 & 100.0 & 100.0 & 100.0 & 100.0 \\
Online Life Sciences \#3 & 1 & 0.0 & 0.0 & 33.0 & 33.0 & 33.0 & 33.0 & 66.0 & 33.0 & 33.0 & 66.0 \\
Online Life Sciences \#4 & 8 & 12.4 & 12.4 & 28.9 & 37.3 & 28.9 & 41.3 & 24.8 & 37.1 & 24.8 & 57.9 \\
Online Life Sciences \#5 & 7 & 61.7 & 57.0 & 66.4 & 80.7 & 71.3 & 71.1 & 61.7 & 66.4 & 61.7 & 85.4 \\
Online Life Sciences \#6 & 2 & 66.5 & 83.0 & 66.5 & 83.0 & 100.0 & 83.0 & 83.0 & 83.0 & 83.0 & 100.0 \\
Online Life Sciences \#7 & 3 & 66.3 & 22.0 & 55.3 & 33.0 & 22.0 & 22.0 & 44.0 & 22.0 & 22.0 & 88.7 \\
\midrule
Online Physics \#1 & 3 & 55.3 & 66.0 & 100.0 & 100.0 & 77.7 & 100.0 & 77.3 & 100.0 & 100.0 & 100.0 \\
Online Physics \#2 & 2 & 66.5 & 49.5 & 83.0 & 33.0 & 66.5 & 100.0 & 83.0 & 83.0 & 66.5 & 100.0 \\
Online Physics \#3 & 7 & 66.4 & 37.9 & 61.4 & 47.3 & 71.0 & 71.0 & 66.3 & 75.9 & 61.6 & 75.9 \\
Online Physics \#4 & 4 & 83.0 & 58.3 & 49.8 & 33.0 & 58.0 & 91.5 & 74.8 & 58.0 & 66.5 & 91.5 \\
Online Physics \#5 & 3 & 66.3 & 22.0 & 55.3 & 55.3 & 66.3 & 66.3 & 55.3 & 55.3 & 55.3 & 77.3 \\
Online Physics \#6 & 13 & 48.4 & 40.8 & 68.8 & 45.8 & 56.0 & 63.7 & 58.6 & 63.7 & 61.2 & 71.4 \\
Online Physics \#7 & 12 & 46.8 & 41.3 & 44.0 & 38.5 & 46.9 & 60.7 & 46.8 & 49.5 & 41.3 & 80.2 \\
Online Physics \#8 & 4 & 58.0 & 41.3 & 49.5 & 66.3 & 41.3 & 49.5 & 49.8 & 58.0 & 49.5 & 74.8 \\
Online Physics \#9 & 7 & 18.9 & 9.4 & 28.3 & 37.7 & 37.7 & 37.7 & 28.3 & 33.0 & 33.0 & 47.1 \\
Online Physics \#10 & 27 & 75.1 & 77.6 & 86.3 & 88.8 & 66.4 & 87.5 & 87.5 & 88.8 & 86.3 & 97.5 \\
\midrule
$^*$Physics \#1 & 24 & 49.8 & 52.5 & 59.4 & 51.0 & 49.7 & 45.5 & 56.6 & 48.3 & 53.8 & 71.9 \\
$^*$Physics \#2 & 45 & 58.2 & 64.9 & 66.3 & 57.5 & 52.2 & 53.7 & 73.1 & 57.4 & 56.0 & 87.2 \\
Physics \#3 & 3 & 33.0 & 33.0 & 33.0 & 33.0 & 55.0 & 33.0 & 44.0 & 44.0 & 33.0 & 66.0 \\
Physics \#4 & 24 & 37.2 & 33.0 & 51.0 & 46.8 & 49.6 & 41.3 & 46.9 & 48.3 & 44.1 & 63.5 \\
$^*$Physics \#5 & 14 & 69.4 & 69.3 & 78.2 & 61.2 & 60.3 & 70.8 & 76.3 & 61.0 & 74.5 & 89.0 \\
$^*$Physics \#6 & 68 & 66.4 & 69.7 & 68.8 & 68.3 & 69.3 & 74.3 & 66.8 & 77.7 & 71.3 & 90.5 \\
$^*$Physics \#7 & 28 & 66.4 & 70.0 & 59.3 & 63.9 & 61.6 & 77.2 & 72.4 & 70.0 & 68.9 & 89.1 \\
$^*$Physics \#8 & 53 & 46.2 & 46.2 & 53.7 & 51.2 & 40.5 & 45.5 & 49.3 & 57.5 & 46.8 & 71.3\\
$^*$Physics \#9 & 478 & 56.8 & 61.2 & 60.4 & 61.2 & 58.0 & 57.3 & 61.8 & 60.8 & 59.0 & 83.2\\
 \bottomrule
 \end{tabular}
 }
 \caption{\textbf{Performance of GPT-4 on open-answer questions for all courses categorized by prompting strategy.} { Majority corresponds to the performance of the majority vote aggregation strategy. Max corresponds to the maximum performance (the score when only one prompting strategy is required to return a correct answer for the model get the answer correct). Online courses typically have fewer open-answer questions as most evaluation in online courses is done through MCQA. $^*$ denotes required courses for a program (applies only for Bachelor and Master programs).}}\label{tab:gpt4_openanswer}
 \end{table*}

\begin{table*}
 \centering
 \fontsize{9.0pt}{\baselineskip}\selectfont % font size
 \renewcommand\tabcolsep{2.5pt} % column space
 \renewcommand\arraystretch{0.85} % row space
  \resizebox{\linewidth}{!}{

 \begin{tabular}{l ccccccccccc}
 \toprule
  \multirow{2}{*}{\textbf{Course name}}& \multirow{2}{*}{\textbf{\# questions}}&\multirow{2}{*}{\textbf{Zero-Shot}} & \multirow{2}{*}{\textbf{One-Shot}} & \textbf{CoT} & \textbf{CoT} & {\textbf{Tree-of-}} & {\textbf{Meta-}} & \multirow{2}{*}{\textbf{Expert}} & \multirow{2}{*}{\textbf{Self-Reflect}} & \multirow{2}{*}{\textbf{Majority}}& \multirow{2}{*}{\textbf{Max}}\\
&&&& (Zero-Shot) & (Four-Shot) & \textbf{Thought}& \textbf{cognitive} & & \\
\midrule
$^*$Biology \#1 & 48 & 62.5 & 75.0 & 81.3 & 83.3 & 81.3 & 79.2 & 56.3 & 83.3 & 85.4 & 87.5 \\
\midrule
Computer Science \#2 & 54 & 31.5 & 35.2 & 46.3 & 42.6 & 35.2 & 35.2 & 27.8 & 38.9 & 42.6 & 59.3 \\
$^*$Computer Science \#4 & 27 & 44.4 & 59.3 & 74.1 & 70.4 & 81.5 & 74.1 & 55.6 & 70.4 & 77.8 & 92.6 \\
$^*$Computer Science \#5 & 20 & 60.0 & 80.0 & 90.0 & 90.0 & 85.0 & 80.0 & 55.0 & 75.0 & 90.0 & 100.0 \\
Computer Science \#8 & 229 & 49.3 & 55.0 & 58.1 & 64.9 & 57.6 & 60.7 & 51.1 & 53.7 & 66.8 & 85.2 \\
$^*$Computer Science \#10 & 158 & 46.8 & 44.9 & 64.3 & 65.6 & 58.9 & 65.8 & 45.6 & 61.1 & 65.8 & 86.7 \\
Computer Science \#11 & 69 & 62.3 & 75.4 & 85.5 & 85.5 & 88.4 & 81.2 & 71.0 & 84.1 & 91.3 & 95.7 \\
$^*$Computer Science \#12 & 36 & 36.1 & 33.3 & 75.0 & 80.6 & 63.9 & 63.9 & 30.6 & 80.6 & 77.8 & 94.4 \\
Computer Science \#13 & 60 & 48.3 & 56.7 & 65.0 & 65.0 & 61.0 & 63.3 & 51.7 & 73.3 & 65.0 & 88.3 \\
$^*$Computer Science \#14 & 111 & 36.0 & 39.6 & 54.1 & 57.7 & 56.8 & 55.9 & 37.8 & 56.8 & 54.1 & 85.6 \\
$^*$Computer Science \#15 & 676 & 56.4 & 67.6 & 78.1 & 79.9 & 77.1 & 77.8 & 58.7 & 76.9 & 82.1 & 94.8 \\
Computer Science \#16 & 41 & 48.8 & 63.4 & 80.5 & 87.8 & 73.2 & 80.5 & 63.4 & 73.2 & 85.4 & 95.1 \\
\midrule
$^*$Math \#1 & 118 & 19.5 & 16.1 & 34.7 & 39.8 & 27.1 & 34.7 & 21.2 & 38.1 & 35.6 & 61.0 \\
$^*$Math \#2 & 31 & 29.0 & 32.3 & 38.7 & 32.3 & 54.8 & 38.7 & 29.0 & 35.5 & 41.9 & 87.1 \\
\midrule
Online Life Sciences \#1 & 286 & 45.5 & 55.6 & 46.9 & 49.3 & 47.9 & 45.8 & 49.5 & 39.2 & 53.5 & 77.6 \\
Online Life Sciences \#2 & 33 & 63.6 & 78.8 & 78.8 & 72.7 & 72.7 & 78.8 & 66.7 & 72.7 & 81.8 & 87.9 \\
Online Life Sciences \#3 & 53 & 32.1 & 26.4 & 37.7 & 34.0 & 43.4 & 47.2 & 34.0 & 35.8 & 47.2 & 75.5 \\
Online Life Sciences \#4 & 226 & 43.4 & 54.9 & 47.8 & 48.2 & 49.6 & 54.4 & 45.6 & 56.6 & 51.3 & 81.4 \\
Online Life Sciences \#5 & 78 & 60.3 & 70.5 & 64.1 & 69.2 & 60.3 & 64.1 & 62.8 & 60.3 & 71.8 & 88.5 \\
Online Life Sciences \#6 & 85 & 54.1 & 62.4 & 60.0 & 61.2 & 58.8 & 57.6 & 58.8 & 57.6 & 60.0 & 82.4 \\
Online Life Sciences \#7 & 48 & 41.7 & 54.2 & 70.8 & 70.8 & 56.3 & 72.9 & 50.0 & 62.5 & 70.8 & 85.4 \\
Online Life Sciences \#8 & 156 & 78.2 & 90.4 & 87.8 & 87.8 & 86.5 & 85.9 & 82.7 & 83.3 & 93.6 & 98.7 \\
\midrule
Online Physics \#1 & 90 & 65.6 & 68.9 & 73.3 & 66.7 & 66.7 & 67.8 & 57.8 & 72.2 & 73.3 & 92.2 \\
Online Physics \#2 & 70 & 61.4 & 72.9 & 71.4 & 77.1 & 75.7 & 74.3 & 58.6 & 74.3 & 78.6 & 94.3 \\
Online Physics \#3 & 74 & 50.0 & 56.8 & 47.3 & 50.0 & 44.6 & 52.7 & 46.6 & 56.8 & 56.8 & 81.1 \\
Online Physics \#4 & 40 & 50.0 & 57.5 & 52.5 & 55.0 & 45.0 & 60.0 & 37.5 & 50.0 & 50.0 & 82.5 \\
Online Physics \#5 & 32 & 37.5 & 56.3 & 68.8 & 53.1 & 68.8 & 56.3 & 34.4 & 65.6 & 68.8 & 93.8 \\
Online Physics \#6 & 55 & 45.5 & 58.2 & 41.8 & 56.4 & 40.0 & 45.5 & 45.5 & 50.9 & 50.9 & 80.0 \\
Online Physics \#7 & 33 & 63.6 & 60.6 & 57.6 & 54.5 & 60.6 & 69.7 & 63.6 & 57.6 & 66.7 & 84.8 \\
Online Physics \#8 & 111 & 48.6 & 61.3 & 61.3 & 57.7 & 61.3 & 65.8 & 54.9 & 61.3 & 61.3 & 84.7 \\
Online Physics \#9 & 60 & 48.3 & 53.3 & 63.3 & 65.0 & 61.7 & 61.7 & 63.3 & 65.0 & 65.0 & 90.0 \\
Online Physics \#10 & 51 & 39.2 & 50.9 & 43.1 & 41.2 & 54.9 & 52.9 & 47.1 & 64.7 & 56.9 & 80.4 \\
Online Physics \#11 & 107 & 59.8 & 70.1 & 69.2 & 71.9 & 69.2 & 67.3 & 52.3 & 60.7 & 74.8 & 86.0 \\
\midrule
Physics course \#3 & 58 & 53.4 & 55.2 & 65.5 & 67.2 & 58.6 & 60.3 & 51.7 & 69.0 & 69.0 & 89.7 \\
Physics course \#4 & 36 & 44.4 & 41.7 & 50.0 & 55.6 & 52.8 & 58.3 & 38.9 & 55.6 & 69.4 & 88.9 \\
 \bottomrule
 \end{tabular}
 }
 \caption{\textbf{Performance of GPT-4 on MCQs for all courses categorized by prompting strategy.} { Majority corresponds to the performance of the majority vote aggregation strategy. Max corresponds to the maximum performance (the score when only one prompting strategy is required to return a correct answer for the model get the answer correct). $^*$ denotes required courses for a program (applies only for Bachelor and Master programs).}}\label{tab:gpt4_mcq}
 \end{table*}

\begin{table*}
 \centering
 \fontsize{9.0pt}{\baselineskip}\selectfont % font size
 \renewcommand\tabcolsep{2.5pt} % column space
 \renewcommand\arraystretch{0.85} % row space
  \resizebox{\linewidth}{!}{

 \begin{tabular}{l ccccccccccc}
 \toprule
  \multirow{2}{*}{\textbf{Course name}}& \multirow{2}{*}{\textbf{\# questions}}&\multirow{2}{*}{\textbf{Zero-Shot}} & \multirow{2}{*}{\textbf{One-Shot}} & \textbf{CoT} & \textbf{CoT} & {\textbf{Tree-of-}} & {\textbf{Meta-}} & \multirow{2}{*}{\textbf{Expert}} & \multirow{2}{*}{\textbf{Self-Reflect}} & \multirow{2}{*}{\textbf{Majority}}& \multirow{2}{*}{\textbf{Max}}\\
&&&& (Zero-Shot) & (Four-Shot) & \textbf{Thought}& \textbf{cognitive} & & \\
\midrule
$^*$Biology \#1 & 12 & 63.4 & 66.3 & 69.2 & 66.3 & 49.5 & 66.3 & 68.9 & 69.1 & 63.4 & 91.5 \\
\midrule
$^*$Chemistry \#1 & 74 & 60.1 & 62.7 & 61.9 & 65.5 & 65.0 & 62.8 & 59.6 & 66.5 & 65.1 & 86.3 \\
\midrule
$^*$Computer Science \#1 & 42 & 46.5 & 53.7 & 59.1 & 52.1 & 46.5 & 52.8 & 53.6 & 56.9 & 52.0 & 84.7 \\
Computer Science \#2 & 98 & 52.1 & 60.9 & 57.5 & 55.4 & 49.3 & 52.1 & 54.5 & 58.2 & 54.5 & 81.1 \\
$^*$Computer Science \#3 & 42 & 59.2 & 48.0 & 60.7 & 63.2 & 55.1 & 58.5 & 65.6 & 60.8 & 55.2 & 83.1 \\
$^*$Computer Science \#4 & 42 & 64.0 & 58.5 & 59.2 & 64.8 & 49.7 & 66.4 & 64.8 & 55.2 & 57.6 & 91.2 \\
$^*$Computer Science \#5 & 223 & 58.2 & 60.6 & 56.2 & 57.6 & 51.3 & 59.5 & 59.5 & 57.7 & 58.9 & 81.8 \\
$^*$Computer Science \#6 & 72 & 40.9 & 43.2 & 41.3 & 35.8 & 37.2 & 38.1 & 40.4 & 39.0 & 39.0 & 62.6 \\
$^*$Computer Science \#7 & 70 & 76.4 & 72.5 & 72.9 & 76.3 & 66.3 & 73.0 & 72.9 & 78.8 & 74.0 & 93.7 \\
Computer Science \#8 & 36 & 58.9 & 54.4 & 55.2 & 57.0 & 34.9 & 46.9 & 57.1 & 48.7 & 51.5 & 75.7 \\
$^*$Computer Science \#9 & 55 & 56.0 & 66.3 & 69.3 & 70.0 & 60.1 & 62.0 & 65.1 & 70.6 & 63.9 & 87.1 \\
$^*$Computer Science \#10 & 34 & 69.4 & 69.3 & 69.3 & 70.4 & 54.6 & 62.5 & 71.3 & 69.3 & 65.4 & 91.1 \\
\midrule
$^*$Data Science \#1 & 28 & 54.4 & 59.2 & 57.9 & 56.8 & 54.4 & 55.6 & 67.5 & 54.5 & 59.2 & 82.0 \\
Data Science \#2 & 36 & 49.6 & 48.7 & 56.1 & 47.7 & 48.8 & 56.1 & 45.9 & 45.9 & 43.1 & 79.3 \\
\midrule
$^*$Math \#1 & 103 & 40.8 & 39.8 & 40.8 & 41.1 & 41.7 & 43.3 & 42.4 & 39.5 & 41.1 & 62.8 \\
$^*$Math \#2 & 302 & 50.4 & 50.4 & 50.4 & 43.9 & 47.2 & 55.7 & 52.3 & 48.9 & 51.9 & 72.6 \\
\midrule
Online Life Sciences \#1 & 9 & 33.1 & 47.8 & 47.7 & 66.4 & 47.9 & 58.9 & 62.6 & 70.1 & 55.2 & 92.6 \\
Online Life Sciences \#2 & 1 & 66.0 & 66.0 & 100.0 & 33.0 & 66.0 & 66.0 & 66.0 & 33.0 & 66.0 & 100.0 \\
Online Life Sciences \#3 & 1 & 0.0 & 0.0 & 33.0 & 33.0 & 33.0 & 33.0 & 33.0 & 33.0 & 33.0 & 33.0 \\
Online Life Sciences \#4 & 8 & 37.3 & 37.3 & 41.3 & 37.1 & 41.4 & 24.8 & 24.8 & 33.0 & 37.1 & 66.4 \\
Online Life Sciences \#5 & 7 & 28.3 & 52.3 & 52.1 & 37.9 & 47.3 & 37.9 & 52.1 & 56.7 & 47.4 & 71.1 \\
Online Life Sciences \#6 & 2 & 100.0 & 100.0 & 100.0 & 100.0 & 66.5 & 83.0 & 83.0 & 100.0 & 100.0 & 100.0 \\
Online Life Sciences \#7 & 3 & 22.0 & 44.3 & 55.3 & 33.0 & 66.3 & 33.3 & 55.3 & 55.3 & 55.3 & 66.3 \\
\midrule
Online Physics \#1 & 3 & 55.3 & 44.0 & 77.7 & 100.0 & 44.0 & 100.0 & 100.0 & 77.7 & 100.0 & 100.0 \\
Online Physics \#2 & 2 & 49.5 & 83.0 & 83.0 & 66.5 & 66.5 & 83.0 & 66.0 & 49.5 & 49.5 & 100.0 \\
Online Physics \#3 & 7 & 52.1 & 52.0 & 52.0 & 47.3 & 56.7 & 33.0 & 37.7 & 47.1 & 42.4 & 71.1 \\
Online Physics \#4 & 4 & 58.0 & 58.0 & 49.8 & 33.0 & 33.0 & 49.5 & 49.5 & 41.5 & 49.8 & 66.3 \\
Online Physics \#5 & 3 & 33.0 & 22.0 & 44.3 & 55.3 & 33.0 & 22.0 & 66.3 & 55.3 & 22.0 & 88.7 \\
Online Physics \#6 & 13 & 43.2 & 38.1 & 50.9 & 43.2 & 40.7 & 45.8 & 58.6 & 40.7 & 45.8 & 68.9 \\
Online Physics \#7 & 12 & 41.3 & 46.9 & 38.5 & 38.5 & 44.0 & 41.3 & 38.6 & 52.4 & 38.5 & 71.8 \\
Online Physics \#8 & 4 & 41.5 & 33.0 & 58.0 & 49.5 & 49.5 & 41.3 & 49.8 & 58.0 & 49.5 & 66.5 \\
Online Physics \#9 & 7 & 33.0 & 18.9 & 37.7 & 28.3 & 33.0 & 18.9 & 37.7 & 42.4 & 37.7 & 51.9 \\
Online Physics \#10 & 27 & 68.9 & 62.8 & 73.9 & 87.6 & 59.0 & 76.3 & 80.0 & 80.1 & 77.7 & 96.2 \\
\midrule
$^*$Physics \#1 & 24 & 30.3 & 38.6 & 33.0 & 37.2 & 23.4 & 33.0 & 31.7 & 28.9 & 35.8 & 51.0 \\
$^*$Physics \#2 & 45 & 43.4 & 47.8 & 39.7 & 42.6 & 36.7 & 36.7 & 41.9 & 39.7 & 37.5 & 65.6 \\
Physics \#3 & 3 & 33.0 & 33.0 & 44.0 & 22.0 & 44.0 & 44.0 & 33.0 & 44.0 & 33.0 & 66.0 \\
Physics \#4 & 24 & 35.9 & 33.0 & 30.3 & 35.8 & 39.9 & 38.5 & 37.2 & 38.5 & 35.8 & 56.5 \\
$^*$Physics \#5 & 14 & 57.1 & 40.0 & 50.0 & 64.2 & 38.6 & 43.4 & 50.1 & 63.8 & 52.0 & 79.7 \\
$^*$Physics \#6 & 68 & 59.4 & 57.6 & 59.0 & 49.1 & 48.6 & 54.5 & 54.6 & 54.6 & 54.5 & 77.7 \\
$^*$Physics \#7 & 28 & 58.1 & 62.8 & 50.9 & 48.5 & 45.0 & 56.9 & 54.5 & 43.7 & 54.5 & 74.7 \\
$^*$Physics \#8 & 53 & 38.6 & 43.0 & 38.6 & 38.0 & 34.9 & 39.9 & 37.4 & 36.8 & 37.4 & 58.1 \\
$^*$Physics \#9 & 478 & 43.2 & 45.5 & 42.5 & 42.5 & 40.0 & 42.7 & 44.6 & 42.6 & 42.0 & 63.7 \\
\bottomrule
\end{tabular}
}
\caption{\textbf{Performance of GPT-3.5 on open-answer questions for all courses categorized by prompting strategy.} { Majority corresponds to the performance of the majority vote aggregation strategy. Max corresponds to the maximum performance (the score when only one prompting strategy is required to return a correct answer for the model get the answer correct). Online courses typically have fewer open-answer questions as most evaluations in online courses are done through MCQA. $^*$ denotes required courses for a program (applies only for Bachelor and Master programs).}}\label{tab:gpt3_openanswer}
\end{table*}

\begin{table*}
 \centering
 \fontsize{9.0pt}{\baselineskip}\selectfont % font size
 \renewcommand\tabcolsep{2.5pt} % column space
 \renewcommand\arraystretch{0.85} % row space
  \resizebox{\linewidth}{!}{

 \begin{tabular}{l ccccccccccc}
 \toprule
  \multirow{2}{*}{\textbf{Course name}}& \multirow{2}{*}{\textbf{\# questions}}&\multirow{2}{*}{\textbf{Zero-Shot}} & \multirow{2}{*}{\textbf{One-Shot}} & \textbf{CoT} & \textbf{CoT} & {\textbf{Tree-of-}} & {\textbf{Meta-}} & \multirow{2}{*}{\textbf{Expert}} & \multirow{2}{*}{\textbf{Self-Reflect}} & \multirow{2}{*}{\textbf{Majority}}& \multirow{2}{*}{\textbf{Max}}\\
&&&& (Zero-Shot) & (Four-Shot) & \textbf{Thought}& \textbf{cognitive} & & \\
\midrule
$^*$Biology \#1 & 48 & 39.6 & 50.0 & 45.8 & 54.2 & 56.2 & 41.7 & 50.0 & 43.8 & 56.2 & 79.2 \\
\midrule
Computer Science \#2 & 54 & 22.2 & 25.9 & 24.1 & 33.3 & 27.8 & 31.5 & 29.6 & 24.1 & 31.5 & 61.1 \\
$^*$Computer Science \#4 & 27 & 37.0 & 48.1 & 37.0 & 44.4 & 51.9 & 44.4 & 48.1 & 48.1 & 55.6 & 77.8 \\
$^*$Computer Science \#5 & 20 & 70.0 & 65.0 & 65.0 & 60.0 & 50.0 & 65.0 & 60.0 & 50.0 & 70.0 & 90.0 \\
Computer Science \#8 & 229 & 47.6 & 48.0 & 38.4 & 43.2 & 33.2 & 43.2 & 46.7 & 36.2 & 52.0 & 88.2 \\
$^*$Computer Science \#10 & 158 & 37.9 & 30.4 & 37.6 & 45.9 & 40.5 & 40.5 & 37.3 & 39.5 & 44.3 & 80.4 \\
Computer Science \#11 & 69 & 59.4 & 63.8 & 63.8 & 69.6 & 53.6 & 65.2 & 62.3 & 60.9 & 71.0 & 89.9 \\
$^*$Computer Science \#12 & 36 & 44.4 & 36.1 & 47.2 & 47.2 & 25.0 & 27.8 & 27.8 & 41.7 & 50.0 & 86.1 \\
Computer Science \#13 & 60 & 25.0 & 28.3 & 38.3 & 50.0 & 33.3 & 46.7 & 31.7 & 43.3 & 45.0 & 76.7 \\
$^*$Computer Science \#14 & 111 & 43.2 & 30.6 & 42.3 & 49.5 & 41.8 & 40.5 & 36.0 & 30.6 & 42.3 & 87.4 \\
$^*$Computer Science \#15 & 676 & 51.3 & 55.6 & 53.4 & 55.9 & 43.8 & 54.9 & 52.9 & 52.2 & 62.1 & 90.1 \\
Computer Science \#16 & 41 & 36.6 & 63.4 & 48.8 & 51.2 & 34.1 & 51.2 & 48.8 & 51.2 & 63.4 & 82.9 \\
\midrule
$^*$Math \#1 & 118 & 14.4 & 14.4 & 15.3 & 19.5 & 9.4 & 18.6 & 13.6 & 10.2 & 19.5 & 38.1 \\
$^*$Math \#2 & 31 & 19.4 & 32.3 & 41.9 & 29.0 & 32.3 & 38.7 & 32.3 & 32.3 & 32.3 & 83.9 \\
\midrule
Online Life Sciences \#1 & 286 & 45.5 & 48.6 & 44.1 & 45.1 & 38.6 & 47.9 & 48.3 & 42.7 & 51.0 & 81.5 \\
Online Life Sciences \#2 & 33 & 69.7 & 66.7 & 78.8 & 60.6 & 78.8 & 66.7 & 69.7 & 69.7 & 78.8 & 93.9 \\
Online Life Sciences \#3 & 53 & 22.6 & 28.3 & 24.5 & 32.1 & 39.6 & 33.9 & 39.6 & 28.3 & 35.8 & 77.4 \\
Online Life Sciences \#4 & 226 & 48.2 & 50.0 & 48.2 & 42.9 & 44.7 & 47.8 & 47.3 & 46.5 & 55.3 & 83.2 \\
Online Life Sciences \#5 & 78 & 58.9 & 61.5 & 66.7 & 52.6 & 47.4 & 61.5 & 55.1 & 57.7 & 69.2 & 84.6 \\
Online Life Sciences \#6 & 85 & 48.2 & 57.6 & 58.8 & 57.6 & 54.1 & 67.1 & 56.5 & 52.9 & 61.2 & 88.2 \\
Online Life Sciences \#7 & 48 & 56.2 & 39.6 & 52.1 & 45.8 & 50.0 & 50.0 & 52.1 & 50.0 & 68.8 & 87.5 \\
Online Life Sciences \#8 & 156 & 66.0 & 75.6 & 73.1 & 69.9 & 63.5 & 69.2 & 69.2 & 60.9 & 80.8 & 93.6 \\
\midrule
Online Physics \#1 & 90 & 52.2 & 52.2 & 55.6 & 52.2 & 51.7 & 53.3 & 51.1 & 52.2 & 56.7 & 84.4 \\
Online Physics \#2 & 70 & 52.9 & 60.0 & 48.6 & 52.9 & 47.1 & 62.9 & 54.3 & 41.4 & 62.9 & 90.0 \\
Online Physics \#3 & 74 & 44.6 & 44.6 & 40.5 & 27.0 & 40.5 & 55.4 & 48.6 & 41.9 & 45.9 & 81.1 \\
Online Physics \#4 & 40 & 42.5 & 32.5 & 40.0 & 45.0 & 27.5 & 50.0 & 42.5 & 41.0 & 52.5 & 77.5 \\
Online Physics \#5 & 32 & 25.0 & 25.0 & 43.8 & 37.5 & 37.5 & 18.8 & 43.8 & 43.8 & 40.6 & 84.4 \\
Online Physics \#6 & 55 & 50.9 & 41.8 & 34.5 & 34.5 & 34.5 & 47.3 & 43.6 & 37.0 & 50.9 & 89.1 \\
Online Physics \#7 & 33 & 51.5 & 42.4 & 56.3 & 51.5 & 39.4 & 63.6 & 57.6 & 51.5 & 60.6 & 87.9 \\
Online Physics \#8 & 111 & 48.6 & 49.5 & 50.5 & 45.9 & 36.0 & 44.1 & 50.5 & 47.7 & 52.3 & 80.2 \\
Online Physics \#9 & 60 & 50.0 & 58.3 & 53.3 & 53.3 & 50.0 & 61.7 & 43.3 & 48.3 & 60.0 & 88.3 \\
Online Physics \#10 & 51 & 52.9 & 47.1 & 49.0 & 43.1 & 41.2 & 43.1 & 43.1 & 49.0 & 58.8 & 76.5 \\
Online Physics \#11 & 107 & 45.8 & 57.9 & 52.3 & 56.1 & 45.8 & 59.8 & 48.6 & 47.7 & 58.9 & 89.7 \\
\midrule
Physics \#3 & 58 & 39.7 & 46.6 & 50.9 & 27.6 & 44.8 & 51.7 & 37.9 & 41.4 & 46.6 & 87.9 \\
Physics \#4 & 36 & 47.2 & 38.9 & 33.3 & 50.0 & 36.1 & 47.2 & 44.4 & 33.3 & 50.0 & 88.9 \\
\bottomrule
 \end{tabular}
 }
 \caption{\textbf{Performance of GPT-3.5 on MCQs for all courses categorized by prompting strategy.} { Majority corresponds to the performance of the majority vote aggregation strategy. Max corresponds to the maximum performance (the score when only one prompting strategy is required to return a correct answer for the model get the answer correct). $^*$ denotes required courses for a program (applies only for Bachelor and Master programs).}}\label{tab:gpt3_mcq}
 \end{table*}

\subsection{Impact of Prompting Strategy}\label{ap:prompting_strategy}
Table~\ref{tab:gpt4_gpt35_prompts} shows the average GPT-4 and GPT-3.5 performance for each prompting strategy. GPT-4 outperforms GPT-3.5 across all prompting strategies.
When answering MCQs, four-shot CoT \citep{wei2023chainofthought} emerges as GPT-4's best-performing strategy, while zero-shot achieves the lowest performance. Curiously, the same ranking does not transfer to open-answer questions, where self-reflect \citep{madaan2023selfrefine} emerges as the best strategy, followed by Expert Prompting \citep{xu2023expertprompting}. Zero-shot prompting remains the least performant. However, based on a survey of reports submitted by Masters students for a class project in a Natural Language Processing (NLP) course, we found students to be most likely to use Zero-shot, Expert, and Zero-shot COT prompting, as these are the most intuitive strategies and the ones that require the least amount of preparation work.

 % \newpage

\begin{table*}[h]
 \centering
 \fontsize{9.0pt}{\baselineskip}\selectfont % font size
 \renewcommand\tabcolsep{2.5pt} % column space
 \renewcommand\arraystretch{0.85} % row space
  % \resizebox{\linewidth}{!}{

 \begin{tabular}{l  l cccccccc}
 \toprule
\multirow{2}{*}{\textbf{Question Type}} & \multirow{2}{*}{\textbf{Model}} & \multirow{2}{*}{\textbf{Zero-Shot}} & \multirow{2}{*}{\textbf{One-Shot}} & \textbf{CoT} & \textbf{CoT} & {\textbf{Tree-of-}} & {\textbf{Meta-}} & \multirow{2}{*}{\textbf{Expert}} & \multirow{2}{*}{\textbf{Self-Reflect}}\\
&&&& (Zero-Shot) & (Four-Shot) & \textbf{Thought}& \textbf{cognitive} & & \\
\midrule
\multirow{2}{*}{MCQ} & GPT-3.5 & $46.4\pm1.7$ & $48.5\pm1.7$ & $47.8\pm1.7$ & $48.3\pm1.7$ & $42.1\pm1.7$ & $\underline{49.8}\pm1.7$ & $47.6\pm1.7$ & $45.0\pm1.7$ \\
& GPT-4 & $50.1\pm1.6$ & $58.7\pm1.6$ & $63.2\pm1.6$ &  $\underline{64.8}\pm1.6$ & $62.1\pm1.6$ &  $63.8\pm1.6$ & $52.2\pm1.6$ & $62.5\pm1.6$ \\
\midrule
\multirow{2}{*}{Open Answer} & GPT-3.5 & $52.7\pm1.4$ & $53.9\pm1.4$ & $53.8\pm1.4$ & $52.5\pm1.4$ & $48.4\pm1.3$ & $53.9\pm1.4$ & $\underline{54.5}\pm1.4$ & $53.5\pm1.4$ \\
& GPT-4 & $62.8\pm1.5$ & $62.9\pm1.5$ & $66.4\pm1.4$ & $64.3\pm1.5$ & $63.9\pm1.4$ & $65.9\pm1.4$ & $67.6\pm1.4$ & $\underline{69.1}\pm1.4$ \\

 \bottomrule
 \end{tabular}
 % % }
 \caption{\textbf{Performance of GPT-3.5 and GPT-4 on all MCQs and open-answer questions, categorized by prompting strategy.} { The most effective prompting strategy for each model is \ul{underlined}. Open-answer questions are graded by GPT-4. For clarity, we have rounded the 95\% confidence intervals for each prompting strategy to one decimal place.}}\label{tab:gpt4_gpt35_prompts}
 \end{table*} 

\subsection{Performance by Language}

In our dataset, we have 70.5\% of English questions and 29.5\% of French questions. Table \ref{tab:language_by_q} shows performance by language across models and question types. Table \ref{tab:gpt4_lang_prompts} shows GPT-4 performance per language per prompting strategy across all question types. 

\begin{table*}[h]
 \centering
 \begin{tabular}{l  l ccc}
 \toprule
\textbf{Question Type} & \textbf{Model} & \textbf{English} & \textbf{French} \\
\midrule
\multirow{2}{*}{MCQ} & GPT-3.5 & $47.3 \pm 1.7$ & $32.5 \pm 4.9$  \\
& GPT-4 & $63.2 \pm 1.7$ & $48.0 \pm 5.1$ \\
\midrule
\multirow{2}{*}{Open Answer} & GPT-3.5 & $52.1 \pm 1.9$ & $54.5 \pm 2.0$ \\
& GPT-4 & $65.4 \pm 2.0$  & $67.3 \pm 2.0$  \\
 \bottomrule
 \end{tabular}
 \caption{\textbf{Performance of GPT-3.5 and GPT-4 on MCQs and open-answer questions, categorized by the question language.} { Open-answer questions are graded by GPT-4. Performance is presented with 95\% confidence interval.}}\label{tab:language_by_q}
 \end{table*} 

\begin{table*}[h]
 \centering
 \fontsize{9.0pt}{\baselineskip}\selectfont % font size
 \renewcommand\tabcolsep{2.5pt} % column space
 \renewcommand\arraystretch{0.85} % row space

 % \resizebox{\linewidth}{!}{
 \begin{tabular}{l  l cccccccc}
 \toprule
\multirow{2}{*}{\textbf{Language}} & \multirow{2}{*}{\textbf{Question Type}} & \multirow{2}{*}{\textbf{Zero-Shot}} & \multirow{2}{*}{\textbf{One-Shot}} & \textbf{CoT} & \textbf{CoT} & {\textbf{Tree-of-}} & {\textbf{Meta-}} & \multirow{2}{*}{\textbf{Expert}} & \multirow{2}{*}{\textbf{Self-Reflect}} \\
&&&& (Zero-Shot) & (Four-Shot) & \textbf{Thought}& \textbf{cognitive} & & \\
\midrule
\vspace{2pt}
\multirow{2}{*}{English} & MCQ & $51.8 \pm 1.7$ & $60.4 \pm 1.7$ & $64.4 \pm 1.7$ & $\underline{66.0} \pm 1.7$ & $63.2 \pm 1.7$ & $65.0 \pm 1.7$ & $53.4 \pm 1.7$ & $63.3 \pm 1.7$ \\
& Open Answer & $62.4 \pm 2.1$ & $62.5 \pm 2.1$ & $67.4 \pm 2.0$ &  $65.7 \pm 2.0$ & $61.8 \pm 2.0$ & $63.2 \pm 2.0$ & $67.7 \pm 2.0$ & $\underline{70.0} \pm 2.0$ \\
\midrule
% \vspace{2pt}
\multirow{2}{*}{French} & MCQ & $37.4 \pm 5.3$ & $42.6 \pm 5.4$ & $51.8 \pm 5.4$ &  $53.3 \pm 5.4$ & $51.9 \pm 5.4$ & $52.3 \pm 5.3$ & $40.7 \pm 5.4$ & $\underline{55.7} \pm 5.4$ \\
& Open Answer & $63.1 \pm 2.1$ & $63.1 \pm 2.1$ & $65.3 \pm 2.0$ & $63.1 \pm 2.1$ & $65.7 \pm 2.0$ & $\underline{68.4} \pm 2.0$ & $67.5 \pm 2.0$ & $\underline{68.4}\pm 2.1$ \\

 \bottomrule
 \end{tabular}
 % }
 \caption{\textbf{Performance of GPT-4 per language on MCQs and open-answer questions, categorized by prompting strategy. }{ The most effective prompting strategy for each language is \ul{underlined}. Open-answer questions are graded by GPT-4. All scores are provided with 95\% confidence intervals.}}\label{tab:gpt4_lang_prompts}
 \end{table*} 

\subsection{Impact of Prompt Language}\label{ap:language}
Tables \ref{tab:language_by_q} and \ref{tab:gpt4_lang_prompts} show differing performance on English questions compared to French questions. Unfortunately, the subsets of courses in our dataset in English and French mostly do not intersect, precluding a conclusive comparison between these performance measurements. However, given that AI assistants are often predominantly trained on English text data, these results raise a question of whether performance on French questions could be increased further, through creative cross-lingual prompting. 

\begin{figure*}[t]
     \centering
     \includegraphics[width=\textwidth]{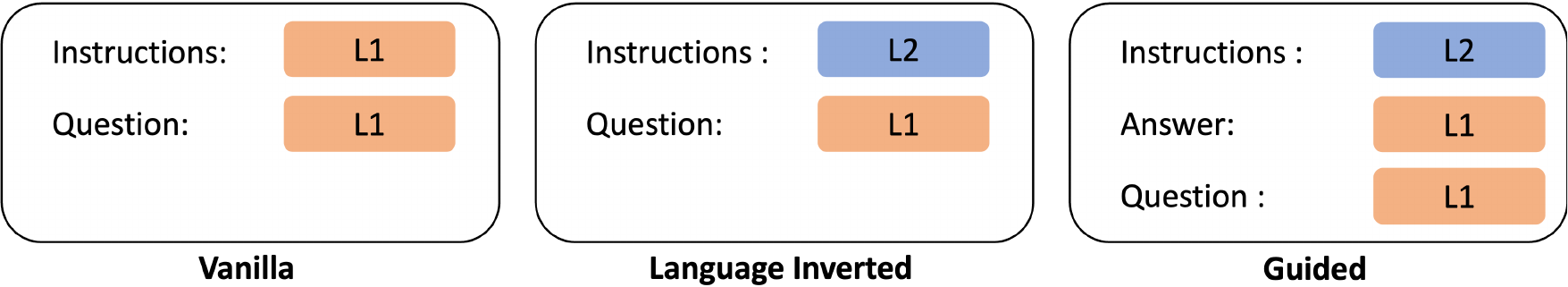}
    \caption{The three language-related prompting strategies. Given two languages \textit{L1} and \textit{L2}, and a question in language \textit{L1}, (1) \textbf{Vanilla}: provides instructions in \textit{L1}; (2) \textbf{Language Inverted}: provides instructions in \textit{L2}; (3) \textbf{Guided}: provides instructions in \textit{L2}, specifying that the question is in \textit{L1}, and that it should be answered in \textit{L1} as well.}
    \label{fig:prompt_language} 
\end{figure*}

Consequently, we explore whether a student user could achieve better performance by varying the language of the prompting instruction. We employ three variations of the metacognitive prompting strategy (which ranks among the top-performing strategies), where we vary the language and the wording of the instruction and the question, as schematized in Figure \ref{fig:prompt_language}: \textit{Vanilla}, \textit{Language-inverted}, and \textit{Guided}. In the \textit{Vanilla} setting, we provide the prompt instruction and question in the same language. In the \textit{Inverted} setting, we provide the instruction and question in different languages. Finally, in the \textit{Guided} setting, we provide the instruction and question in different languages but clarify in the instruction that the answer should be provided in the same language as the question. We focus on MCQ-based performance to avoid potential language bias from GPT-4 as a grader, assessing the impact of these three variations across all English and French MCQs of our dataset.

\begin{table*}[h]
\centering
% \resizebox{\linewidth}{!}{
\begin{tabular}{l cc cc cc}
\toprule
 \multirow{2}{*}{\textbf{\shortstack[l]{Question\\Language}}}  & \multicolumn{2}{c}{\textbf{Vanilla}} & \multicolumn{2}{c}{\textbf{Inverted}} & \multicolumn{2}{c}{\textbf{Guided}} \\ 
 &  GPT-3.5 & GPT-4 & GPT-3.5 & GPT-4 & GPT-3.5 & GPT-4 \\

\midrule

 English &  $51.2 \pm 1.8 $ &$ \underline{65.4} \pm 1.5 $ & $44.5 \pm 1.7 $ & $61.7 \pm 1.6 $ & $46.8 \pm 1.7 $ & $61.4 \pm 1.6 $ \\
% \midrule

French & $39.4 \pm 4.2 $ & $\underline{53.1} \pm 4.3 $ & $38.3 \pm 4.1 $ & $51.9 \pm 4.4 $ & $36.2 \pm 4.0 $ & $53.0 \pm 4.1 $ \\

\bottomrule
\end{tabular}
% }
\caption{\textbf{Performance comparison of GPT-3.5 and GPT-4 across the three different prompting strategies} (\textit{vanilla}, \textit{language inverted}, and \textit{guided}), categorized by question language. All scores are provided with 95\% confidence interval.}\label{tab:language_metacog_scores}
\end{table*}

As illustrated in Table \ref{tab:language_metacog_scores}, the average scores for the \textit{Vanilla} setting are higher for both English and French compared to the \textit{language-inverted} setting, indicating that instructing the model in the same language as the question leads to higher performance for both models compared to when the question is in a different language. Finally, guiding the model by asking it to reason and answer in the same language as the question, even if the instructions are in another language (i.e., the \textit{guided} setting), enhances the performance for GPT-4 on French questions, yielding a score equivalent to providing instructions in French. Taken together, our results show that there is little benefit from prompting the model in English (a language that most pretrained models have likely seen more data from) compared to the language of the question.

\end{document}